\documentclass[useAMS,usenatbib]{mn2e}
\usepackage{rotating}
\usepackage{subfigure}
\RequirePackage{setspace}
\usepackage{graphicx}
\usepackage{fancyhdr}
\usepackage{natbib}
\usepackage{amsmath}
\usepackage{amssymb}
\usepackage{aas_macros}
\usepackage{epstopdf}
\usepackage{lscape}
\usepackage{rotating}
\usepackage{lineno}

\title[GAMA: SFR in interacting pairs]{Galaxy And Mass Assembly (GAMA): the effect of close interactions on star formation in galaxies}
\author[L. J. M. Davies et. al.]{L. J. M. Davies$^{1}$\thanks{E-mail: luke.j.davies@uwa.edu.au}, A. S. G. Robotham$^{1}$, S. P. Driver$^{1,2}$,  M. Alpaslan$^{3}$,  I. K. Baldry$^{4}$, \newauthor J. Bland-Hawthorn$^{5}$, S. Brough$^{6}$,  M. J. I. Brown$^{8}$,  M. E. Cluver$^{10}$,  M. J. Drinkwater$^{11}$,  \newauthor C. Foster$^{6}$, M. W. Grootes$^{7}$,  I. S. Konstantopoulos$^{6}$,   M. A. Lara-L\'opez$^{9}$, \newauthor   \'A. R. L\'opez-S\'anchez$^{6,12}$,  J. Loveday$^{13}$, M. J. Meyer$^{1}$, A. J. Moffett$^{1}$,    P. Norberg$^{14}$,  \newauthor  M. S. Owers$^{6,12}$, C. C. Popescu $^{15,16}$,  R. De Propris$^{17}$,   R. Sharp$^{18}$,  R. J. Tuffs$^{7}$, \newauthor L. Wang$^{14,22}$, S. M. Wilkins$^{13}$, L. Dunne$^{19,20}$, N. Bourne$^{20}$, M. W. L. Smith$^{21}$ \\
 \\
$^{1}$ ICRAR, The University of Western Australia, 35 Stirling Highway, Crawley, WA 6009, Australia \\
$^{2}$ SUPA, School of Physics and Astronomy, University of St Andrews, North Haugh, St Andrews, Fife, KY16 9SS, UK\\
$^{3}$ NASA Ames Research Centre, N232, Moffett Field, Mountain View, CA 94035, United States of America \\
$^{4}$ Astrophysics Research Institute, Liverpool John Moores University, IC2, Liverpool Science Park, 146 Brownlow Hill, Liverpool, L3 5RF, UK\\
$^{5}$ Sydney Institute for Astronomy, School of Physics A28, University of Sydney, NSW 2006, Australia \\
$^{6}$ Australian Astronomical Observatory, PO Box 915, North Ryde, NSW 1670, Australia\\
$^{7}$ Max Planck Institut fuer Kernphysik, Saupfercheckweg 1, 69117 Heidelberg, Germany \\
$^{8}$ School of Physics and Astronomy, Monash University, Clayton, Victoria 3800, Australia \\
$^{9}$ Instituto de Astronom\'ia, Universidad Nacional Aut\'onoma de M\'exico, A.P. 70-264, 04510 M\'exico, D.F., M\'exico\\
$^{10}$ University of the Western Cape, Robert Sobukwe Road, Bellville, Cape Town 7535, South Africa\\
$^{11}$ School of Mathematics and Physics, University of Queensland, Brisbane, QLD 4072, Australia \\
$^{12}$ Department of Physics and Astronomy, Macquarie University, NSW 2109, Australia\\
$^{13}$ Astronomy Centre, Department of Physics and Astronomy, University of Sussex, Falmer, Brighton BN1 9QH\\
$^{14}$ Institute for Computational Cosmology, Department of Physics, Durham University, Durham, DH1 3LE, UK\\
$^{15}$ Jeremiah Horrocks Institute, University of Central Lancashire, PR1 2HE Preston, UK\\
$^{16}$ The Astronomical Institute of the Romanian Academy, Str. Cutitul de Argint 5, Bucharest, Romania\\
$^{17}$ Finnish Centre for Astronomy with ESO, University of Turku, Vais\"{a}l\"{a}ntie 20, Piikki\"{o}, 21500, Finland\\
$^{18}$ Research School of Astronomy \& Astrophysics, Mount Stromlo Observatory, Cotter Road, Weston Creek, ACT 2611, Australia\\
$^{19}$ Department of Physics and Astronomy, University of Canterbury, Private Bag 4800, Christchurch, 8140, New Zealand\\
$^{20}$ Institute for Astronomy, University of Edinburgh, Royal Observatory, Edinburgh EH9 3HJ, UK\\
$^{21}$ School of Physics \& Astronomy, Cardiff University, The Parade, Cardiff CF24 3AA, UK. \\
$^{22}$ SRON Netherlands Institute for Space Research, Landleven 12, 9747 AD, Groningen, The Netherlands.}

\begin{document}
\date{Accepted: May 2015}
\pagerange{\pageref{firstpage}--\pageref{lastpage}} \pubyear{2015}
\maketitle

\begin{abstract}
The modification of star formation (SF) in galaxy interactions is a complex process, with SF observed to be both enhanced in major mergers and suppressed in minor pair interactions. Such changes likely to arise on short timescales and be directly related to the galaxy-galaxy interaction time. Here we investigate the link between dynamical phase and direct measures of SF on different timescales for pair galaxies, targeting numerous star-formation rate (SFR) indicators and comparing to pair separation, individual galaxy mass and pair mass ratio. We split our sample into the higher (primary) and lower (secondary) mass galaxies in each pair and find that SF is indeed enhanced in all primary galaxies but suppressed in secondaries of minor mergers. We find that changes in SF of primaries is consistent in both major and minor mergers, suggesting that SF in the more massive galaxy is agnostic to pair mass ratio.

We also find that SF is enhanced/suppressed more strongly for short-time duration SFR indicators ($e.g.$ H$\alpha$), highlighting recent changes to SF in these galaxies, which are likely to be induced by the interaction. We propose a scenario where the lower mass galaxy has its SF suppressed by gas heating or stripping, while the higher mass galaxy has its SF enhanced, potentially by tidal gas turbulence and shocks. This is consistent with the seemingly contradictory observations for both SF suppression and enhancement in close pairs.

\end{abstract}


\section{Introduction}
\label{sec:Intro}

Interactions are a key process in the evolution of galaxies in the Universe \citep[$e.g.$][]{White91}. As galaxies interact and merge hierarchically over cosmic timescales, these interaction events leave strong imprints on the galaxies involved, modifying their morphology \citep[$e.g.$][, or see review in Conselice (2014)]{Conselice03,Conselice05,Lotz08a, Lotz08b, Mortlock13}, triggering AGN \citep[$e.g.$][however $c.f.$ \citealp{Villforth14}]{Kazantzidis05,Hopkins08,Medling13}, varying gas fractions \citep[$e.g.$][]{Kazantzidis05,Lotz10,Ueda12,Ueda13} and significantly altering their star-formation (SF) history \citep[$e.g.$][]{DiMatteo07, Shi09, Wong11, Bournaud11, Patton13, Robotham13}. Such merger events are ubiquitous throughout the Universe and as such, probing the details of the merger process is key to our understanding of galaxy formation and evolution \cite[$e.g.$][]{Casteels14}. For example, star-formation (SF) initiated by interactions could well be a key factor in transforming blue star-forming discs into red, passively evolving quiescent spheroids.     

However, the details of how SF responds to galaxy interactions are somewhat vague and are likely to be strongly linked to pair mass ratio and pair separation. Traditionally, observational studies of close pairs have highlighted strong evidence that SF is enhanced through interactions and that the strongest enhancement occurs in the closest pairs \citep[those at projected separations of $<$30\,kpc, $e.g.$][]{Ellison08,Freeman10}. In addition, detailed multi-wavelength analyses of blue compact dwarf galaxies have shown that, in the majority of the cases, interactions with or between low-luminosity dwarf galaxies or HI clouds are the triggering mechanism of their strong star-formation activity \citep{Lopez-Sanchez10, Lopez-Sanchez12}. More recently, \citet{Patton13} have studied a large sample of star-forming galaxies in pairs taken from the Sloan Digital Sky Survey (SDSS) and find clear evidence for enhanced, emission-line derived  SF out to significant pair separations ($\sim150$\,kpc). They also find that the enhancement in SF is inversely proportional to pair separation, with the closest pairs displaying the largest enhancement. This result is echoed by \cite{Wong11} who perform a similar analysis for isolated pair galaxies in the PRIsm MUlti-object Survey (PRIMUS) but determine star-formation rates (SFRs) using attenuation corrected Ultraviolet(UV)-optical colours. They find pairs at separations of $<50$\,kpc show bluer far-ultraviolet minus r-band (FUV-$r$) colours than a control sample of non-pair galaxies (indicative of larger SFRs). Moreover, this SFR enhancement is more pronounced at closest pair separations ($<30$\,kpc). \cite{Scudder12} perform a similar analysis on SDSS pairs and, in addition to SF enhancement, find that SF is most strongly enhanced in major merger systems, hinting that pair mass ratio is significant in the modification of SF history through galaxy interactions. In a distinct but complementary approach, \cite{Owers07} find that 2df Galaxy Redshift Survey (2dfGRS) selected star-busts are more likely to have a neighbour of comparable brightness within 20\,kpc, and 30\% of star-bust galaxies show morphological signatures of interactions and mergers - suggesting that the SF in these systems is induced by a galaxy-galaxy interaction.   However, by contrast \citet[]{Li08} also find strong enhancement of SF in interactions but suggest that there is little correlation between this enhancement and the relative luminosity of the interacting galaxies (and therefore, pair mass ratio). They do also find that the enhancement of SF is stronger for lower mass systems. By splitting individual interacting galaxies into high ($<log[M^{*}/M_{\odot}]>=10.6$) and low ($<log[M^{*}/M_{\odot}]>=9.72$) stellar mass samples, they find that at a given pair separation, SF is enhanced more strongly in low mass than high mass pair galaxies. 

Numerical simulations offer further insight into the complexities of these interaction SF processes. \cite{DiMatteo07} model several hundred galaxy close pair interactions ($<20$\,kpc) for various morphological classes, and find that while SF is primarily enhanced in galaxy interactions, mergers do not always trigger starbursts, and galaxy interactions are not always sufficient to convert high gas masses into new stars. They also highlight that the amount by which SF is enhanced is anticorrelated with pair separation on small scales ($i.e.$ galaxies which have a very close passage produce the lowest bursts of SF), as well as the amplitude of tidal forces ($i.e.$ pairs that undergo less intense tidal forces can preserve higher gas masses for future SF during the merger). These simulations suggest that SF enhancement is not ubiquitous in interactions and that SF is likely to vary as a function of pair separation and pair mass ratio - higher mass ratios will induce more significant tidal effects in the lower mass galaxy in the pair removing the bulk of its gas and as such, starving SF. 

Recently we have seen tentative observational evidence in support of these simulation predictions for close pairs. \cite{Robotham13} studied L$^{*}$ galaxies in closely interacting pairs taken from the Galaxy And Mass Assembly (GAMA) Galaxy Group Catalogue (G$^{3}$C), and investigate variations in SF as a function of pair mass ratio. They find that primary (higher mass) pair galaxies have their SF enhanced, but only when the pair mass ratio is close to 1. While the secondary (lower mass) galaxy has its SF suppressed relative to equivalent stellar mass non-pair galaxies. In addition, \cite{DePropris14} find that close pairs galaxies are consistently redder than a similarly selected comparison sample, and suggest that these systems have been ``harassed'' in multiple previous passes prior to the current close interaction. These results give the first tentative observational evidence that SF in interacting systems is more complicated than the traditional evidence would suggest, and highlights that pair mass ratio may play a vital role in the changes to SF induced by an interaction. 

However, questions still remain regarding the details of SF in interacting systems, such as how does SF proceed in interactions as a function of pair separation when considering pair samples split by merger ratio, primary or secondary status and individual galaxy mass? A potentially more significant question is over what timescales do variations in the SFR in interacting systems become apparent? If we measure SFRs using observational tracers which probe different SF timescales, will we see differences in observed SFRs? SF enhancement/suppression in interactions is likely to occur on short timescales, and hence, may only be apparent in SFR measures which probe short periods of the galaxy's SF history.                 

In this paper, we further investigate SF in closely interacting pairs taken from the G$^{3}$C catalogue. We split our sample on merger ratio, primary or secondary status and galaxy mass, and investigate SF characteristics as a function of pair separation. We determine SFRs for each galaxy using multiple nebular emission line and continuum fluxes, which probe the galaxy's SF history over different timescales, and compare these methods in order to identify short timescale variations in SF, once again as a function of pair separation. In this manner we identify factors driving the modification of star-formation rates in interacting galaxies. In Section \ref{sec:data} we outline the GAMA survey and pair catalogue used in this work, Section \ref{sec:SFR_measures} discusses the SFR indicators used, in Section \ref{sec:SF_pairs} we highlight the effects of interactions on SF in galaxies and in Section \ref{sec:summary2} we summarise our results.  

Throughout this paper we use a standard $\Lambda$CDM cosmology with {H}$_{0}$\,=\,70\,kms$^{-1}$\,Mpc$^{-1}$, $\Omega_{\Lambda}$\,=\,0.7 and $\Omega_{M}$\,=\,0.3.

\section{Data}
\label{sec:data}

\subsection{GAMA and the Pair Catalogue}
\label{sec:gama}

The GAMA survey is a highly complete multi-wavelength database \citep{Driver11} and galaxy redshift ($z$) survey \citep{Baldry10,Hopkins13, Liske14} covering 280\,deg$^{2}$ to a main survey limit of $r_{\mathrm{AB}}<19.8$\,mag in three equatorial (G09, G12 and G15) and two southern (G02 and G23) regions. The spectroscopic survey was undertaken using the AAOmega fibre-fed spectrograph \citep[][]{Sharp06,Saunders04} in conjunction with the Two-degree Field \citep[2dF,][]{Lewis02} positioner on the Anglo-Australian telescope and obtained redshifts for $\sim$250,000 targets covering $0<z\lesssim0.5$ with a median redshift of $z\sim0.2$, and highly uniform spatial completeness \citep{Baldry10,Robotham10,Driver11}. Full details of the GAMA survey can be found in \citet{Driver11} and \citet{Liske14}. In this work we utilise the first 5 years of data obtained and frozen for internal team use, referred to as GAMA II.  

\begin{figure}
\begin{center}

\includegraphics[scale=0.58]{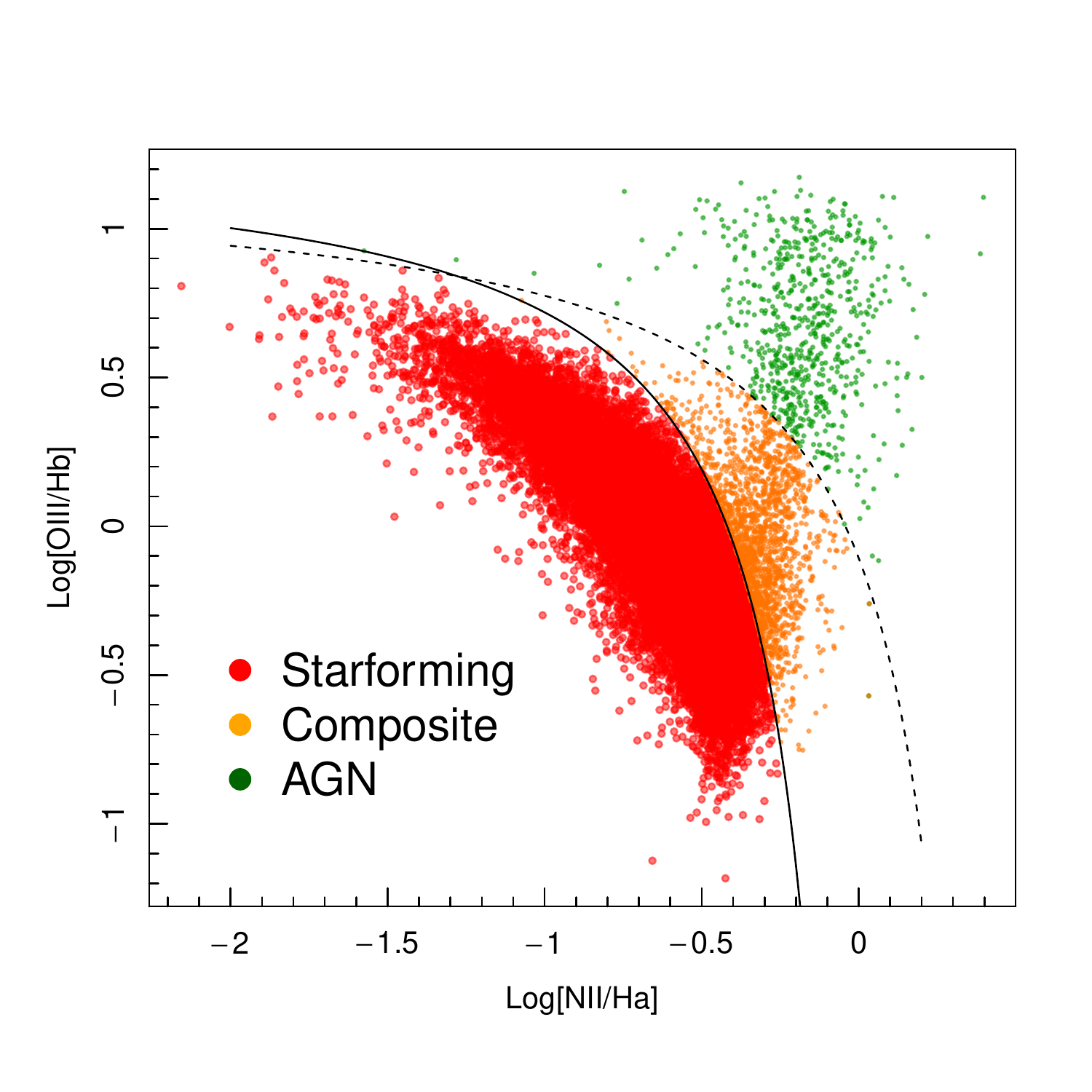}

\caption{The BPT diagram for all GAMA II sources with all diagnostic emission lines detected at S/N $>$ 3. The AGN dividing line of \citet{Kewley13} is displayed as the black dashed line while the composite dividing line of \citet{Kauffmann03} is shown as the black solid line. Sources which potentially contain AGN are displayed as green or orange.}
\label{fig:BPT}
\end{center}
\end{figure}

Firstly, to minimise Active Galactic Nuclei (AGN) contamination in our sample, which may potentially bias our SFR estimates, we exclude optically bright AGN using the BPT diagnostic diagram \citep[][]{Bladwin81}.  We select sources from the GAMA II spectral line catalogue which have all BPT diagnostic lines detected at S/N $>$ 3, and have emission-lines which do not lie at the edge of the spectral range (and may have poor line measurements). We then apply the BPT AGN+composite source selection line of \citet{Kauffmann03}:

\begin{equation}
\mathrm{log_{10}([OIII]/H\beta) > \frac{0.61}{log_{10}([NII]/H\alpha)-0.05} + 1.3},
\end{equation}

\noindent to identify galaxies which may have their SFR contaminated by AGN (Fig. \ref{fig:BPT}). Galaxies which meet the BPT AGN selection are coloured green and those which are potentially composite AGN+SF sources are coloured orange. We find a total of 2,486 galaxies from the full GAMA II sample which potentially contain an optically bright AGN.  

In order to select systems which are undergoing a close interaction, we utilise the GAMA G$^{3}$C catalogue which includes the identification of all galaxy pairs \citep[][also see \citealp{Robotham12,Robotham13,Robotham14}]{Robotham11}. Briefly, pairs are selected on physical projected separation, $r_{\mathrm{sep-proj}}$ (for the cosmology given in Section \ref{sec:Intro}) and radial velocity separation, $v_{\mathrm{sep-rad}}$. In this paper we use systems which meet the pair criteria of \cite{Robotham12,Robotham13,Robotham14}:   

\begin{equation*}
P_{r100v500} = r_{\mathrm{sep-proj}} < 100\,\mathrm{kpc} \wedge v_{\mathrm{sep-rad}} < 500\,\mathrm{km\,s}^{-1}.
\end{equation*}

This sample consists of a total of 37,679 galaxies in pairs. We further restrict our sample to pairs at $z<0.3$ in order to minimise the impact of redshift evolution in our galaxies and biases towards the identification of high mass pairs at the high redshift end, giving a total of 33,832 pair galaxies. We then match all pair galaxies to the GAMA II panchromatic photometry catalogue (Driver et al in prep) and use rest-frame photometry derived from a refactored implementation of the InterRest algorithm \citep{Rudnick03, Taylor09}, coupled with the empirical set of galaxy template spectra of \cite{Brown14}. We exclude any galaxy classed as an AGN using the classifiers described above, leaving 32,468 pair galaxies. 

We use the stellar masses derived from the $ugriZJH$ photometry for all GAMA II galaxies \citep{Taylor11} to assign classes to each individual galaxy. Firstly, we determine whether each galaxy is the primary or secondary system in the interacting pair (by mass). Secondly, we calculate the pair mass ratio and identify each system as being either part of a major merger (mass ratio $<$3:1) or minor merger (mass ratio $>$3:1). Note that we also split our minor merger sample further into 3:1$<$mass ratio$<$6:1 and mass ratio$>$6:1 subsamples, but see little difference between each of these populations. As such, we do not discuss them in this work.

A potential caveat to the selection above, is that we may be biased towards pair systems where both galaxies are star-forming, specifically at the low mass and/or high redshift end - where we are not mass complete. For example, by purely applying a redshift selection we are more sensitive to higher stellar mass and, potentially star-forming, galaxies. As such, low mass systems will only be identified as being in an interaction if both pair galaxies are high mass and/or highly star-forming. Likewise, if we were to only apply a mass selection over the full sample, we do not take into account the broad redshift range of our selection and any evolution of the star-forming properties of galaxies across this extensive look-back time. As such, our pair sample will be biased to wards specific galaxy populations and we will dilute any observed trends by not taking into account the global change in star-formation properties of galaxies with redshift. In order to remove this affect, we compare all of our pair galaxy samples to a mass and redshift matched control sample. The control sample contains the same mass bias and redshift evolution as the pairs sample, and as such removes any dependancy on this effect (GAMA is essentially 100\% complete in the $r$-band, thus sources will also not drop out of our pairs selection preferentially to our control sample due to the effect of the interaction on SF - where the $r$-band will not be strongly affected). As such, we can directly compare the effects of an interaction to non-interacting galaxies at the same stellar mass and redshift. Our control sample is defined further in Sec. \ref{sec:Independent}. In this paper we only display our results in comparison to this control sample, and as such are not biased by our mass incompleteness.   

We note that in defining our control sample based on stellar mass and redshift, we are not taking into account the different galaxy clustering statistics of passive and active systems. Passive galaxies are more strongly clustered \citep[e.g.][]{Zehavi11} and as such, more likely to be in pairs \citep[see][for the fraction of quiescent galaxies in major mergers at $z\sim0$]{Lin08, deRavel09}. This could potentially bias our control sample towards a higher fraction of actively star-forming systems. However, we do not wish to perform any sample selection based on star-formation diagnostics, as the increased fraction of passive systems in pair galaxies may in fact be caused by the galaxy interactions - the very the effect we wish to measure here. As we will see, the majority of sub-samples in our analysis show enhanced star-formation in close pairs, a result that would only be reduced by the potential biases discussed above. In addition, we see distinct differences between the primary and secondary galaxies of minor mergers in our sample. Any bias produced by different clustering statistics would affect both populations in the same manner. Hence, such a bias is unlikely to be driving this result.

In all of the subsequent analysis in this paper, we consider each pair galaxy member individually, but split our samples into primary and secondary galaxies, and major and minor mergers based on the classification within their pair system. We also further split our samples by the stellar mass of each individual galaxy as noted above. For reference, Table \ref{tab:numbers} shows the number of galaxies in each mass bin as a function of primary or secondary status and major or minor merger classification.

\begin{table}
\caption{Number of pair sources (each pair galaxy is treated individually) in each mass bin as a function of primary or secondary system status and major or minor merger classification.}
\begin{center}
\begin{scriptsize}
\begin{tabular}{ |c|c|c|c|c|c}
Mass Range & Total & Primary & Secondary & Major & Minor \\
(log$_{10}$[M$_{\odot}$]) & & & & & \\
\hline
\hline

9.5-10.0 & 4,606 & 1,407 & 3,199 & 2,088 & 2,517 \\ 
10.0-10.5 & 8,795 & 3,958 & 4,837 & 5,121 & 3,670 \\
10.5-11.0 & 10,910 & 7,282 & 3,628 & 6,870 & 4,040\\
\hline
All & 24,311 &12,647 & 11,664 & 14,079 & 10,227 \\
\hline

\end{tabular}
\end{scriptsize}
\end{center}
\label{tab:numbers}
\end{table}

\section{SF measures over different timescales}
\label{sec:SFR_measures}

\begin{figure}
\begin{center}

\includegraphics[scale=0.45]{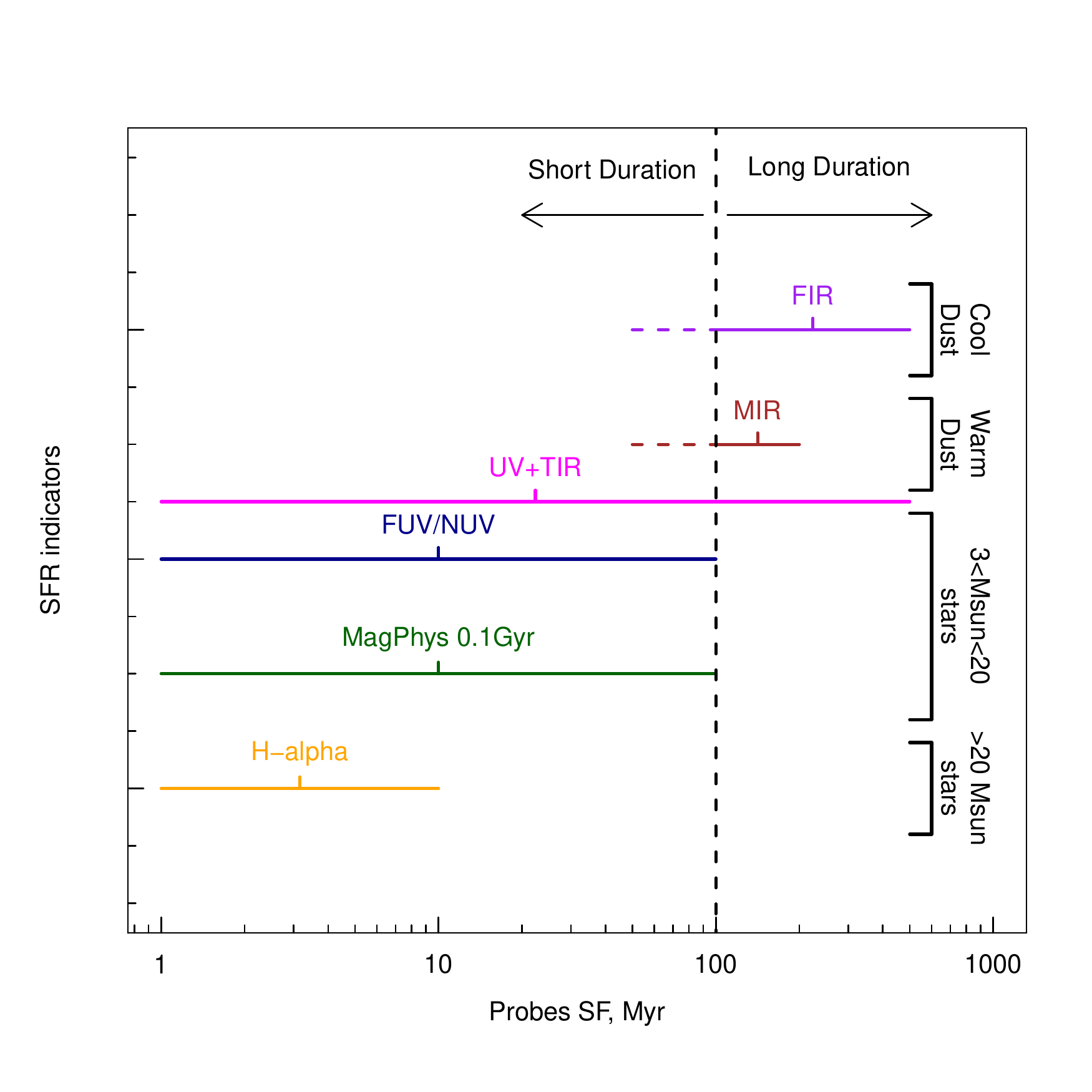}

\caption{SF timescale probed by each SFR indicator used in this work, for normal disk galaxies. Sources of the emission are noted on the right hand side of the figure. We split our measures into short ($<100$\,Myr) and long ($>100$\,Myr) duration. See text for details of each SFR indicator. }
\label{fig:timescales}
\end{center}
\end{figure}

In this paper we consider SFRs determined via different observables in order to probe SF on different timescales and equate them to recent variations in the galaxy's instantaneous SFR. Fig. \ref{fig:timescales} shows a cartoon pictorial representation of the timescales over which each SFR indicator (described in more detail below) probes in normal disk galaxies, and the source of the emission in each case. This figure is only intended to highlight the key differences in timescales over which each indicator probes, as the complex physics of each process is likely to place large variation on the true timescale of the emission arising from each source. As such, we do not constrain our SFR indicators further than splitting them into long duration - probing timescales $>$100\,Myr, far-infrared (FIR), UV+total infrared (UV+TIR) and mid-infrared (MIR), and short duration - SFR indicators probing only $<100$\,Myr, FUV, near-ultraviolet (NUV), MAGPHYS (0.1\,Gyr) and H$\alpha$. See \cite{Gilbank10} for a comparison of different SFR indicators in the SDSS stripe 82 and details of each SFR indicator used in this work below.   

\subsection{Long duration MIR/FIR Continuum SFRs}

Infrared (IR) emission at 1-1000$\mu$m arising from `normal' star-forming galaxies is produced from three sources: photospheres and circumstellar envelopes of old stars undergoing mass loss \citep[$e.g.$][]{Melbourne12}, interstellar gas and dust heated by either bright OB stars in star-forming regions (warm dust) or the general stellar radiation field throughout the Inter Stellar Medium (ISM, cool dust - `cirrus') - for example see the review by \cite{Sauvage05}, \cite{Popescu00} or more recently \cite{Xilouris12} and references therein. Stellar sources of IR emission dominate at short wavelengths $<3\mu$m and interstellar gas emission makes up just a few percent of the total IR output of galaxies. At $3 - 100\mu$m the bulk of the emission arises from warm dust locally heated by UV emission from young stars in star-forming regions, with cirrus emission dominating at $>60\mu$m. As such, probing IR flux from star-forming galaxies in the $3 - 100\mu$m range gives a reliable estimate for the ongoing SF \citep[$e.g.$][- the amount of flux emitted in the IR is directly related to the UV emission from newly formed stars]{Calzetti07}. However, IR emission from dust requires significantly long timescales to become apparent and subsides slowly when SF is suppressed \citep[see][]{Kennicutt98}. As such, SFRs derived from the IR continuum probe SF over large timescales, giving an estimate of a galaxy's SF history on timescales of $\gtrsim100$\,Myr.

In this paper we consider two IR continuum measures of SFR. Firstly, we determine the FIR SFRs derived from the 100$\mu$m flux provided for all GAMA sources as part of the  Herschel Astrophysical Terahertz Large Area Survey \citep[H-ATLAS][]{Eales10} and outlined in the GAMA II panchromatic data release (Driver et al in prep). =The 100$\mu$m data comes from the H-ATLAS Phase 1 Data Release (Valiante et al., in prep), which provides PACS maps at 100 and 160$\mu$m reduced using Scanamorphos. The GAMA-II photometry in the Herschel maps exploits an algorithm developed by \cite{Bourne12} to optimally capture extended flux while accounting for blending in the low-resolution images. This algorithm consists of convolving the 100$\mu$m map with a kernel given by the GAMA $r$-band-defined aperture smoothed with the 9$^{\prime\prime}$ (FWHM) PSF of the PACS 100$\mu$m band. These smoothed apertures are created for all GAMA galaxies, and any that overlap in the map are down-weighted so that the 100$\mu$m flux in such pixels is shared between the overlapping apertures. Fluxes in all other H-ATLAS bands are used for the UV+TIR and Magphys SFRs (see below) and are produced in a similar manner, but using different PSFs - which are specific to the band in question. For more details see \cite{Bourne12} and Driver et al (in prep). We convert 100$\mu$m fluxes to SFRs using the tight correlation derived in \cite{Davies14} for Virgo cluster galaxies:

\begin{equation}
\mathrm{log_{10}\,SFR_{FIR}(M_{\sun}yr^{-1}) = 0.73\,log_{10}L_{100\mu m}(W\,Hz^{-1})- 17.1 }.    
\end{equation}

Secondly, we derive MIR SFRs using the Wide-field Infrared Explorer ($WISE$) data from the matched GAMA-$WISE$ catalogue outlined in \cite{Cluver14}. We use $WISE$ 22$\mu$m (W4) band fluxes, which are not strongly affected by PAH emission, and the best-fit SFR correlation obtained in \cite{Cluver14}:    

\begin{equation}
\mathrm{log_{10} \, SFR_{\mathrm{MIR}} (M_{\odot}\,yr^{-1})=0.84\,\mathrm{log}_{10} \nu L_{22\mu m}(L_{\odot})-7.3}.
\end{equation}

We note that the correlations derived in \cite{Cluver14} compare WISE fluxes with H$\alpha$ derived SFRs and as such, may be biased towards emission line-derived SFRs. However, we rescale all SFRs derived in the paper to have the same slope and normalisation as the SFR$_{FIR}$ in order to compare each measure directly (see below).

\subsection{Short duration attenuation corrected UV Continuum and Magphys SFRs}

In contrast with the FIR and MIR, UV continuum derived SFRs probe a shorter timescale than those in the IR ($\lesssim$ 100\,Myr). UV continuum arises from hot, massive (M$_{*} > 3$M$_{\odot}$) O and B stars, and as such is a good tracer of more recent SF in galaxies \citep[$e.g.$][]{Kennicutt12}, with luminosity-weighted mean ages for a constant SFR predicted to be $\sim$28\,Myr for the GALEX FUV band, and $\sim$80\,Myr for GALEX NUV (Grootes et al. in prep).

Hence, by comparing IR continuum and UV continuum SFRs we may be able to disentangle recent SF changes in a system - potentially due to interactions. 

However, direct estimates of the SFR from UV continuum luminosities are problematic as the UV emission arising from a galaxy is extremely sensitive to dust attenuation \citep[$e.g.$][]{Wang96}. Previous studies have attempted to overcome this issue through estimating attenuation corrected SFRs by applying attenuation estimates based on the UV spectral slope ($\beta$) and luminosity corrections, such as those derived by \cite{Meurer99} - $e.g.$ \cite{Wijesinghe11}. Such corrections apply general scaling to all galaxies and may not be appropriate for specific galaxy classes \citep[$e.g.$][]{Wilkins12}, severely biasing any results derived by such an analysis. 

Here we take a different approach and use the full Spectral Energy Distribution (SED) Magphys \citep{daCunha08} fits to the GAMA II galaxies (Driver et al in prep). In this upcoming work, Driver et al  obtain the best SED fit to the full 21-band photometric data available to all GAMA II sources, simultaneously fitting the UV through FIR flux and obtaining the best-fit un-attenuated model spectrum. For our attenuation corrected UV luminosities we use the unattenuated UV emission from the best-fit Magphys model template convolved with the GALEX FUV/NUV filter curves. In this manner we use each galaxy's full SED to estimate the attenuation correction to be applied to the UV luminosity for each individual galaxy instead of applying a general $\beta$ correction.  We convert UV luminosity to at SFR using the calibrations given in \cite{Wijesinghe11}:

\begin{equation}
\mathrm{SFR_{NUV}(M_{\odot}\,yr^{-1}) = \frac{L_{NUV} (W\,Hz^{-1})}{1.56 \times 10^{21}}},   
\end{equation}
 \begin{equation}
\mathrm{SFR_{FUV}(M_{\odot}\,yr^{-1}) = \frac{L_{FUV} (W\,Hz^{-1})}{1.64 \times 10^{21}}} .   
\end{equation}

The Magphys code also provides an estimate of the galaxy SFR averaged over the last 100\,Myr using a best fit model for both the FIR and UV emission for a combined estimate of both obscured and unobscured SF (hereafter SFR$_{0.1\,Gyr}$). Hence, this provides an additional short timescale SFR estimate with which to compare to our long duration SFR$_{FIR}$.

\subsection{Long duration UV Continuum + Total Infrared Luminosity SFR}

We also use the combination of UV and total IR (TIR) luminosities as a star formation rate proxy broadly probing the last $\sim300$\,Myr of the galaxies star-formation history. As discussed above, UV emission arises directly from star-forming regions and probes SF on short timescales, while some fraction of this emission is absorbed and reprocessed by dust, being re-emitted in the FIR on longer timescales. As such, using a star-formation rate indicator which probes both the UV and FIR emission with give a relatively stable, but broad timescale, measure of star-formation in our sample galaxies.  We therefore sum both UV and TIR luminosities to obtain a total star formation rate estimate, based on the bolometric luminosity of OB stars. 

We use the method outlined  many high redshift studies \citep[e.g.][]{Bell05,Papovich07, Barro11} of:

\begin{equation}
\mathrm{SFR_{UV+TIR}(M_{\odot}\,yr^{-1}) = 1.09 \times 10^{-10}[L_{IR}+ 2.2 L_{UV}](L_{\odot}) }.
\end{equation}

This prescription is the \cite{Bell05} recalibration of the relation from \cite{Kennicutt98}, scaled for a \cite{Chabrier03} stellar IMF. Here, $L_\mathrm{IR}$ is the total IR luminosity, integrated between 8-1000 $\mu$m.  These values have been estimated by fitting the \cite{Elbaz01} galaxy templates to the WISE and Herschel photometric points.  The bolometric UV luminosity between 1216-3000 \AA, $L_\mathrm{UV}$, is estimated as $1.5 \nu f_{\nu, 2800}$, where $f_{\nu,2800}$ is the rest-frame luminosity at 2800 \AA. For further details of this process see Taylor et al. (in prep).

\subsection{Short duration nebular H$\alpha$ SFRs}

H$\alpha$ photons arise from gas ionised by the stellar radiation field, and only stars with ages $<$20\,Myr can contribute significantly to this ionizing flux. Thus, H$\alpha$ provides a direct measure of the current SFR in galaxies ($<$10-20\,Myr) which is largely independent of SF history \citep[$e.g.$ see][]{Kennicutt98}. 

For SFR$_{\mathrm{H\alpha}}$ we use emission line data from the GAMA II spectroscopic campaign (Owers et al. in prep), where aperture, obscuration and stellar absorption corrected H$\alpha$ luminosities are given by:                 

\begin{equation}
\begin{split}
\mathrm{L_{H\alpha}} &= \mathrm{(EW_{H\alpha} + EW_{c})} \times 10^{-0.4(M_{r}-34.1)} \\
                       & \ \ \ \  \mathrm{\times \frac{3\times10^{18}}{(6564.1(1+z)^2)} \left(\frac{F_{H\alpha}/F_{H\beta}}{2.86} \right) ^{2.36}},
\end{split}
\end{equation}

and $\mathrm{EW_{H\alpha}} $ denotes the H$\alpha$ equivalent width, $\mathrm{EW_c}$ is the equivalent width correction for stellar absorption \citep[2.5\AA\ for GAMA,][]{Hopkins13}, $M_{r}$ is the galaxy $r$-band magnitude and $\mathrm{F_{H\alpha}/F_{H\beta}}$ is the Balmer decrement \citep[see][for further details]{Gunawardhana11}. Using this, SFR$_{\mathrm{H\alpha}}$ can be determined from \cite{Kennicutt98}, assuming a Salpeter IMF:

\begin{equation}
\mathrm{SFR_{\mathrm{H\alpha}} (M_{\odot}/yr)=\frac{L_{H\alpha} (W\,Hz^{-1})}{1.27 \times 10^{34} }}.
\end{equation}

One caveat to using H$\alpha$ SFRs is that the aperture based spectroscopy only probes the central regions of nearby galaxies. However, in this paper we only investigate star-formation rates in comparison to a control sample of galaxies matched on mass and redshift (see below) and as such, both our pair and control sample should suffer the same aperture bias.   

We note that the GAMA stellar masses, UV+TIR and Magphys-based SFRs are calculated assuming a Chabrier IMF, while our H$\alpha$, MIR, and FIR SFRs are calculated using a Salpeter IMF, as such we scale all Salpeter IMF SFRs by a factor of 1.5 to account for this discrepancy \citep{Dave08, Driver13}. In addition, all results in this work are displayed relative to a control sample with SFRs derived in an identical manner. As such, differences been IMF assumptions in in different indicators will not affect our results.

\begin{figure*}
\begin{center}

\includegraphics[scale=0.63]{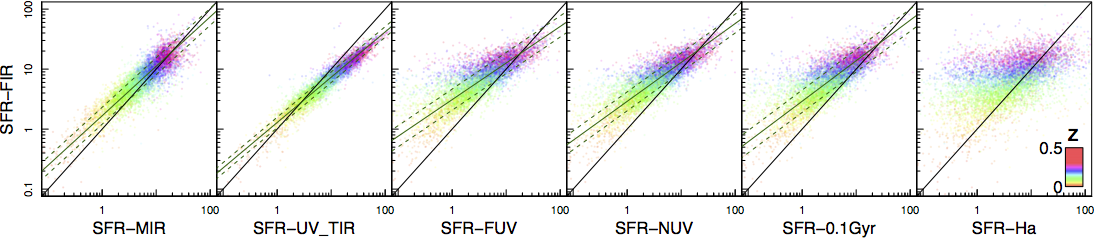}

\caption{Comparisons between various SFR measures used in this work. The 1:1 relation is shown by the solid black line. Points are colour coded by source redshift, z. \textsc{Hyperfit} best fit lines to each correlation are displayed as the green lines with $\pm 1\sigma$ errors on the normalisation. These fits are used to scale each distribution to the same slope and normalisation using equation \ref{eq:sc}. Note we do not scale SFR$_{\mathrm{H\alpha}}$ as the correlation between SFRs is poor. }
\label{fig:SFR_comp}
\end{center}
\end{figure*}

\subsection{Calibrating continuum SFRs}

As noted above, we recalibrate all continuum SFR measures to have the same slope and normalisation as SFR$_{FIR}$. We apply this correction as all SFR indicators are derived using vastly different calibrations, and different samples. If SFR indicators consistently derived SFRs in the same manner, we would find a similar slope and normalisation when comparing indicators.  In Fig. \ref{fig:SFR_comp} we display our SFR indicators in comparison to SFR$_{FIR}$ for all sources detected at S/N $>$ 2 in both the Herschel 100$\mu$m and $WISE$ W4 observations (the limiting bands in our selection). Clearly, not all indicators show the same slope and normalisation. These offsets are likely to be produced by the SF calibration, not the true galaxy population. As such, we fit the slope and normalisation of each distribution using the [R] multi-dimensional MCMC fitting procedure, \textsc{Hyperfit} (Robotham \& Obreschkow, 2015) and recalibrate each SFR measure using the best fit parameter as follows:

 \begin{equation} 
\label{eq:sc}
\mathrm{log_{10} \, SFR_{\mathrm{cal}} = a \times log_{10} \,SFR_{\mathrm{orig}} + b} \\
 \end{equation}

 \noindent where,

 \begin{equation*} 
\mathrm{SFR_{MIR}, a=0.824 \wedge b= 0.223}\\
 \end{equation*} 
  \begin{equation*} 
\mathrm{SFR_{UV+TIR}, a=0.755 \wedge b= 0.110}\\
 \end{equation*}
   \begin{equation*} 
\mathrm{SFR_{FUV}, a=0.610 \wedge b= 0.491}\\
 \end{equation*}  
  \begin{equation*} 
\mathrm{SFR_{NUV}, a=0.649 \wedge b= 0.462}\\
 \end{equation*} 
  \begin{equation*} 
\mathrm{SFR_{0.1\,Gyr}, a=0.595 \wedge b=0.562 }\\
 \end{equation*} 
  \begin{equation*} 
\mathrm{SFR_{H\alpha}, a=1 \wedge  b=0}  
 \end{equation*} 
 
Note that we do not apply a correction to the H$\alpha$ emission line-derived SFR as there is large scatter in the distribution but the locus follows a 1:1 correlation. In fact, this scatter could  be due to the very timescale variations we hope to identify in this work - short time scale variations will be most apparent through a comparison of H$\alpha$ and FIR derived SFRs. We explore this further in Section \ref{sec:scatter}. While we apply these calibrations to form the main comparison in this work, we note that our conclusions hold true if no correction is applied. A comparison of SFR indicators for the full GAMA II sample will be the subject of an upcoming paper (Davies et al in prep). 

\begin{figure}
\begin{center}

\includegraphics[scale=0.6]{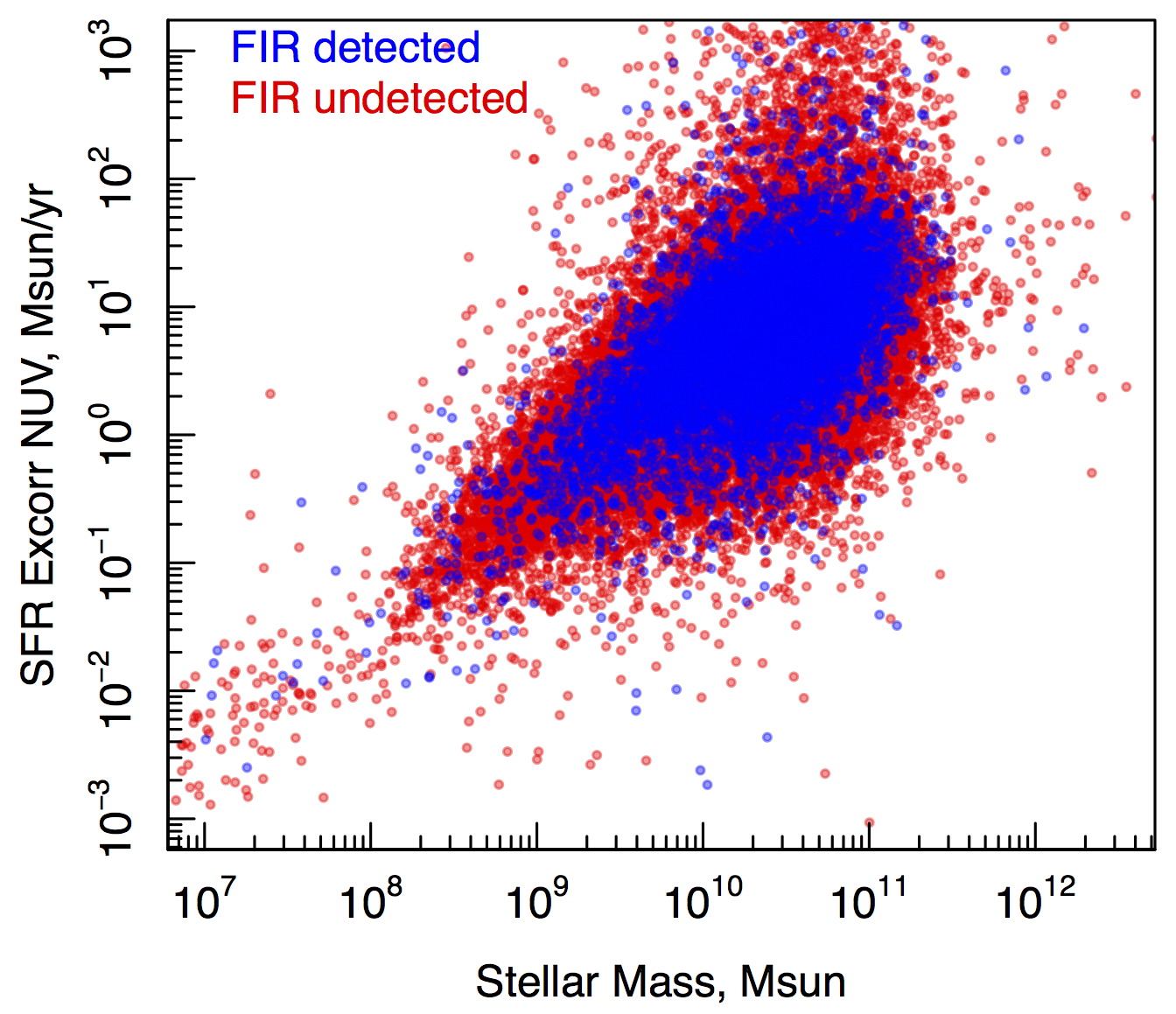}

\caption{The main sequence of star-forming galaxies for extinction corrected NUV SFRs (see text for details). FIR detected sources are shown as blue points, while FIR undetected sources are shown as red points. Both populations are consistent and as such, correlations derived for the FIR detected sources are applicable to the full sample.}
\label{fig:main_sequence}
\end{center}
\end{figure}

One potential caveat to these scalings is that they are defined only for FIR detected sources, and may not be applicable to the general galaxy population. To highlight this this is unlikely to cause significant bias in our results and that these scalings are appropriate, Fig. \ref{fig:main_sequence} shows the main sequence of star-forming galaxies for a completely independent SFR indicator not used in this work. Here we use the extinction corrected NUV SFR (using the direct NUV fluxes and UV-spectral slope extinction correction, unlike the Magphys NUV fits used elsewhere in this work), which does not use any information from the FIR (this SFR indicator will be defined further in Davies et al in prep). For all GAMA II galaxies in our redshift and mass limited sample, we display sources with FIR detections ($>2\sigma$) as blue points, and those without FIR detections as red points. Clearly, the main sequence is consistent between FIR detected and undetected sources, suggesting that any correlations derived from the FIR detected sample is applicable to all galaxies.

In the following analysis, unless otherwise stated, we use specific SFRs (sSFR) calculated by dividing all measured SFRs by stellar mass. This removes any mass dependance on observed SFR correlations.

\section{SF in pair galaxies as a function of pair separation}
\label{sec:SF_pairs}

In attempting to pin down the effect of close interactions on galaxy SFRs we have two measurements which can both be considered as a proxy for the stage of interaction which we are witnessing: i) pair separation and ii) SFRs on different timescales. In the former, to first order, we can equate the pair separation to stage of the interaction - in that distant pairs are more likely to be at an early stage and close pairs at a later stage of the interaction (we will discuss the caveats to this assumption in Section \ref{sec:summary2}). Therefore, we may expect galaxies at large pair separations to display SFRs which are not strongly affected by the interaction and close pairs to show the strongest effects. In the latter, we are directly measuring SFRs at different stages of the interaction. As such, we may expect long duration SFR indicators to probe an epoch prior to the modification of SF, while short duration indicators should once again show the strongest affect. 

In the following three subsections we separate out the variations with pair separation alone, considering each SFR indicator independently, and do not discuss the use of multiple indicators to highlight short duration timescale changes. However, when reading the following analysis we encourage the reader to keep in mind that pair separation is a proxy for timescale within the interaction. The following sections discuss the fine details of SFR variation in pairs and we direct the casual reader to Section \ref{sec:summary1} for a summary of the key results found in the following Section, and Section \ref{sec:summary2} for a full summary of all of our results.

\subsection{Comparisons between pair samples in Robotham et al. (2014)}

\begin{figure}
\begin{center}

\includegraphics[scale=0.45]{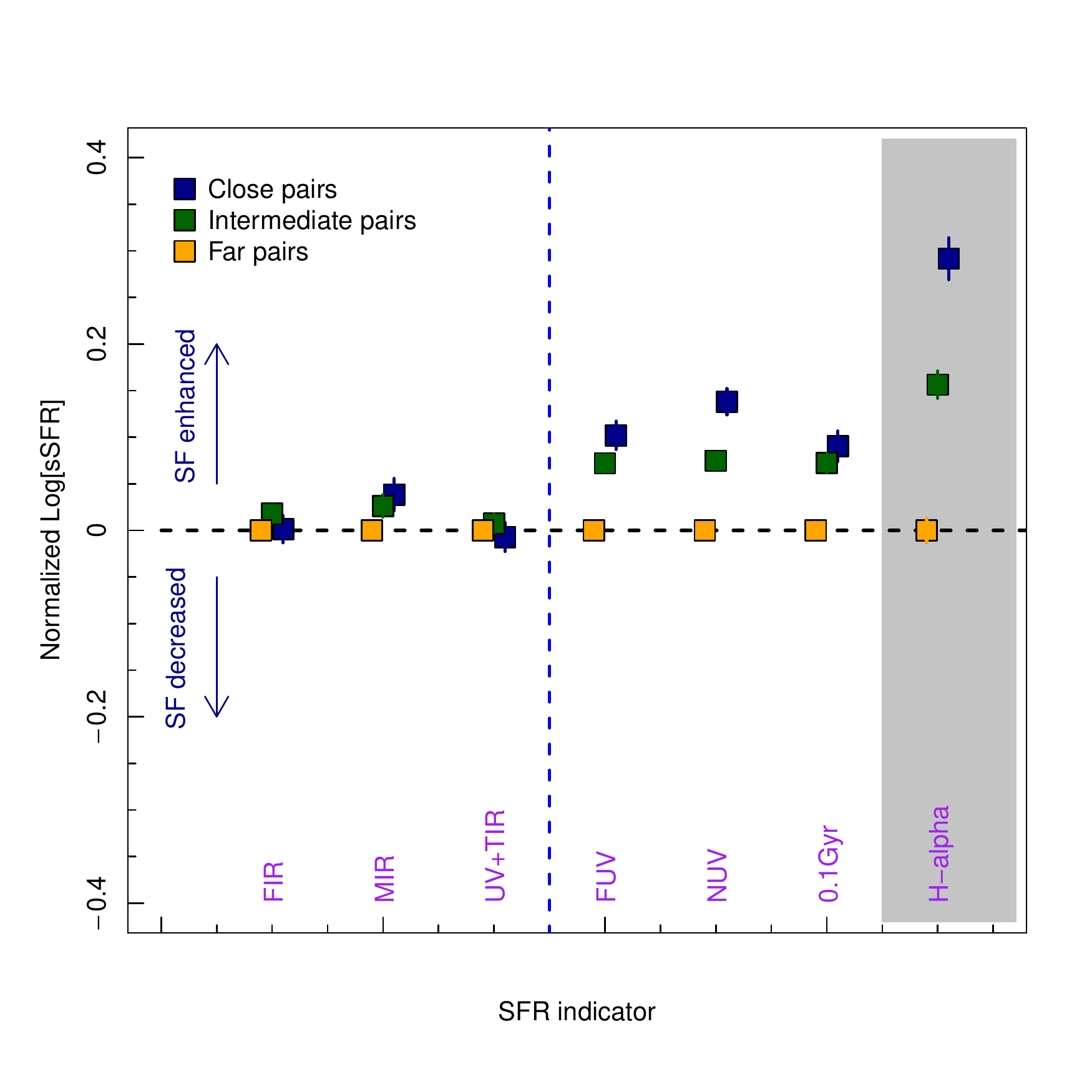}

\caption{Comparison of SFR indicators for pair galaxy samples discussed in \citet{Robotham14}. Points display the median sSFR for each indicator, while errors show the standard error on the median. All points are normalised to the far-pairs value to highlight differences between far-, likely non-interacting, and close-, interacting, pairs. SFR indicators are ordered from long to short duration (left to right). SF appears either constant or increased for close pairs in comparison to far pairs depending on the SFR indicator used.}
\label{fig:aaron_class}
\end{center}
\end{figure}

Initially, we compare different SFR indicators for our pair sample as a whole (not split on mass, pair mass ratio, etc) in order to highlight discrepancies in the results derived from previous studies probing SF in interacting systems. We split our pair systems on the three pair selection criteria outlined in \citet{Robotham14}:

\begin{equation} 
P_{r20v500} = r_{sep} < 20\,h^{-1} \mathrm{kpc} \wedge v_{sep} < 500\,\mathrm{km\,s^{-1}},
 \end{equation}     
\begin{equation*} 
P_{r50v500} = r_{sep} < 50\,h^{-1}\mathrm{kpc}\wedge v_{sep} < 500\,\mathrm{km\,s^{-1}},
 \end{equation*}  
 \begin{equation*} 
P_{r100v1000} = r_{sep} < 100\,h^{-1}\mathrm{kpc} \wedge v_{sep} < 1000\,\mathrm{km\,s^{-1}}
 \end{equation*}  
 
\noindent with each sample excluding sources in the inner contained samples ($i.e.$ galaxies in the P$_{r20v500}$ sample are not contained in the P$_{r50v500}$ or P$_{r100v1000}$ sample). Hereafter we define these samples as close-, intermediate-, and far-pairs respectively. We calculate the median log$_{10}$[sSFR] for each indicator and normalise to the value for far-pairs - to highlight variations in sSFR between closely interacting and distant systems. Fig. \ref{fig:aaron_class} shows the difference between pair samples for each SFR indicator, points are offset for each pair sample in the x-direction for clarity. Error bars display the standard error on the median for each sample, where for many points this is smaller than the plotted symbol. In normalising to the far-pair systems, positive values in Fig. \ref{fig:aaron_class} indicate that SF is enhanced relative to the far-pair sample, while negative values display that it has been suppressed. We find that when measuring SFRs in the FIR, MIR and using UV+TIR SFRs we would see either no change or in fact suppression of sSFR with pair separation, while for other SFR indicators (FUV, NUV, 0.1Gyr and Ha) we would see net enhancement of SF. The most significant variation is seen in the SFR$_{H\alpha}$, where sSFRs appear strongly enhanced by an interaction for both the intermediate- and close-pair samples. This echoes the findings of \citet[][]{Patton13} and others who find strong enhancements in emission line derived SFRs.  

While the results discussed above are seemingly in conflict (or at least highlight that care must be taken when comparing different SFR indicators) they do not consider the vastly different properties of both the individual galaxies and the merger as a whole. In the following subsections we shall investigate this further by splitting our samples on stellar mass, primary(higher mass)/secondary(lower mass) status, and pair mass ratio, and considering variations in SF as a function of pair separation directly (rather than binning into close, intermediate and far pairs). For completeness, in Appendix A we display the samples defined in \citet{Robotham14} but split further into major/minor mergers and primary/secondary status.      

\subsection{Independent measures of SF as a function of Pair Separation, Pair Mass Ratio and Pair Status}
\label{sec:Independent}

In this section, we split our pair galaxy sample into stellar mass bins of $\Delta$log$_{10}$[M*]$=0.5$ from $10^{9.5}$\,M$_{\odot}$ to $10^{11.0}$\,M$_{\odot}$. We do not include galaxies outside of this mass range, where sample sizes are too low to apply the analysis discussed in this work. We are currently investigating star-formation in M$^{*}<10^{9.5}$\,M$_{\odot}$ pair galaxies using a different, but complementary, approach - this will be the subject a future work (Davies et al. in prep).

We then bin on $\Delta$ r$_{\mathrm{sep}}$=10 kpc\,h$^{-1}$ scales ($i.e.$ ignoring line of sight separations which provide little information inside of the initial pair selections). 

As discussed previously, we calculate the excess SF for each indicator in comparison to a mass and redshift matched control sample. In order to define this control sample, for each pair galaxy in our sample we select three corresponding non-pair GAMA galaxies, which are the closest match in stellar mass and redshift parameter space (determined by the closest match in 2D space in terms of 3$\sigma$ clipped standard deviation in each parameter). We do this in each stellar mass, pair mass ratio, primary/secondary status and pair separation bin, such that each date point is scaled appropriately for its matched control sample.  

For each sample, mass range and radial separation bin we then calculate the 3$\sigma$ clipped mean sSFR for both pair and control galaxies. We define the SFR$_{\mathrm{excess}}$ as the ratio of the mean log[sSFR] in pair galaxies, divided by the mean log[sSFR] of non-pair galaxies, as follows:

\begin{equation}
\mathrm{log_{10}[SFR_{ex}(M^{*}, r_{sep})] = \dfrac{\mu(log_{10}[sSFR_{pairs}(M^{*}, r_{sep})])}{\mu(log_{10}[sSFR_{control}(M^{*}, r_{sep})])}.}
\end{equation}

Note that the results discussed in this paper are not sensitive to choice of choice mean/median and all trends are observable when using median distributions over 3$\sigma$ clipped means. We do not show the median distributions here for sake of clarity. Table \ref{tab:numbers2} displays the number of sources used in each sample.

\begin{table}
\caption{Number of pair galaxies used in our analysis for each SFR indicator }
\begin{center}
\begin{scriptsize}
\begin{tabular}{c c}
Indicator & Number of pair galaxies \\
\hline 
\hline

FIR & 12,168 \\
MIR & 4,520 \\
UV+TIR & 10,678 \\
FUV & 24,285 \\
NUV &  24,285 \\
0.1Gyr & 24,285 \\
Ha & 20,170 \\
\hline

\end{tabular}
\end{scriptsize}
\end{center}
\label{tab:numbers2}
\end{table}

The left columns of Figures \ref{fig:100micron} - \ref{fig:Ha} display SFR$_{\mathrm{excess}}$ for our full sample of pair galaxies as a function of pair separation split on stellar mass and into primary and secondary galaxy within the pair (green and orange lines respectively). The coloured polygons show the standard error on the mean for the pairs and non-pairs summed in quadrature at each radial distance. In the middle and right columns of these figures we split our samples further into major and minor mergers respectively. 

A potential pitfall of this analysis is that poor resolution of continuum observations used in this work mean that in many cases close systems share the same resolution element. While attempts have been made to deblend fluxes into individual sources, this may lead to systematic errors in our result. For instance, the secondary galaxy in a pair may have its emission systematically boosted/suppressed via deblending with the primary galaxy and vice versa. In the following figures we display the physical size of the PSF FWHM of the primary instrument used in calculating the star-formation rate, scaled to the median redshift of our pairs sample, as a dashed blue vertical line. As such, sources within this line may have uncertainties on their SFRs which arise from source confusion. However, we note that if real, such uncertainties in this region should be mirrored between primary and secondary populations, $i.e.$ If secondary galaxies systematically have their fluxes underestimated at close pair separations, then primary galaxies should systematically have their fluxes enhanced, as total flux must be distributed between the two sources via deblending. This effect is also likely to be minimal in the Magphys results which use the full SED to estimate SFRs. We also note that this effect will not be apparent in the H$\alpha$ measurements, where the fibre aperture size is much smaller then the physical separation between galaxies. In Sec. \ref{sec:compare} we compare the slopes of these distributions. However, we note here that we consider slopes for both the full range of pair separations, and only at separations larger then the blue line (and so not affected by source confusion) and the results are consistent, albeit at lower significance.

Below we highlight key observables from Figures \ref{fig:100micron} - \ref{fig:Ha}, these will be discussed further in the following sections:

\begin{figure*}
\begin{center}

\includegraphics[scale=0.6]{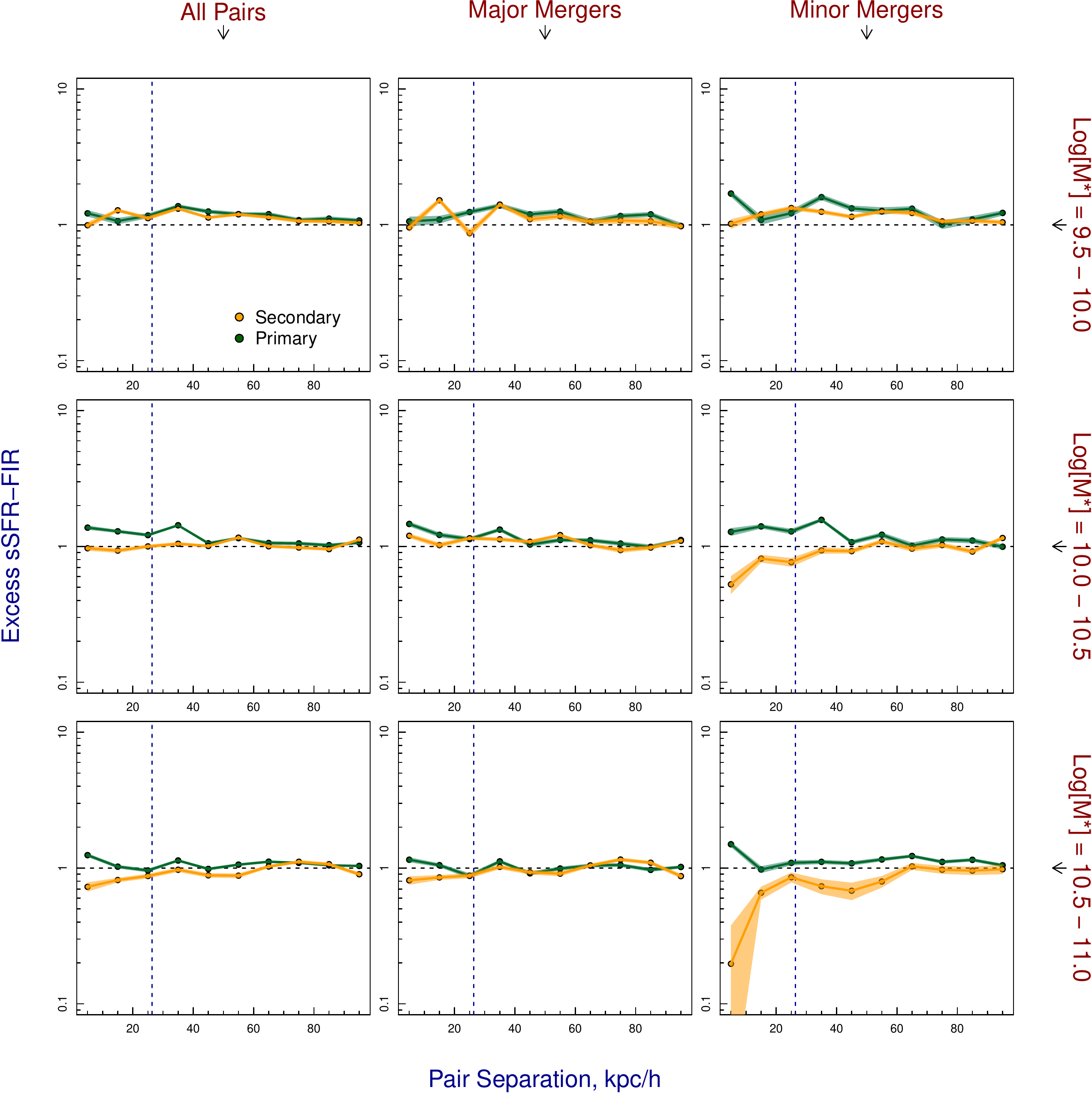}

\caption{The 3$\sigma$ clipped mean excess log$_{10}$[sSFR$_{\mathrm{FIR}}$] for pair galaxies as a function of pair separation binned on $\Delta$ r$_{\mathrm{sep}}$=10 kpc\,h$^{-1}$ scales. Coloured polygons show the standard error on the mean at each separation. We split out pair samples by stellar mass (rows), into major and minor mergers (separated at 3:1 pair mass ratio - columns) and by primary or secondary status within the pair (green and orange lines respectively). Blue dashed vertical line shows the FWHM of the PSF of the PACS 100$\mu$m instrument at the median redshift of our pairs samples.}
\label{fig:100micron}
\end{center}
\end{figure*}

\begin{figure*}
\begin{center}

\includegraphics[scale=0.6]{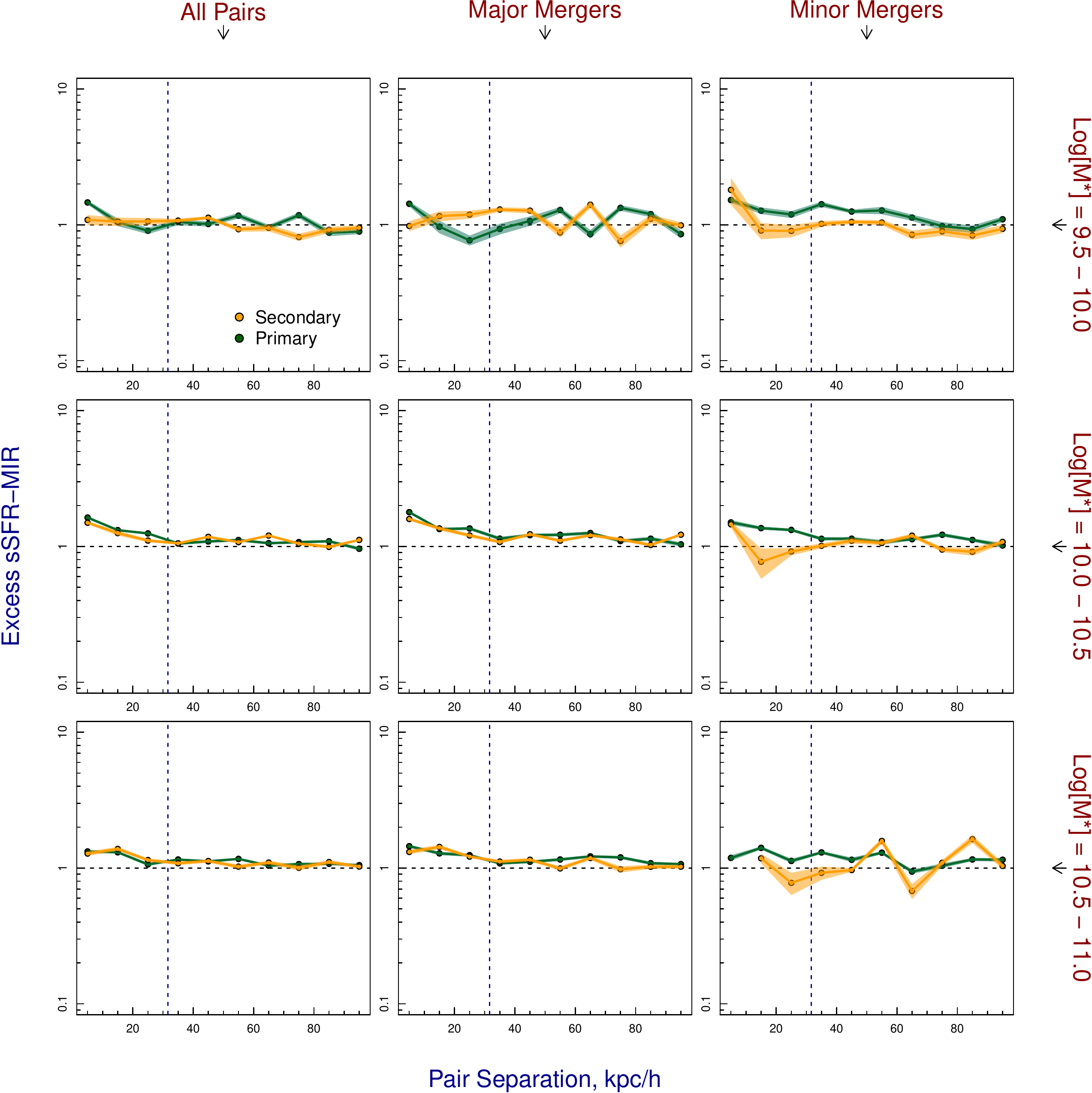}

\caption{Same as Fig. \ref{fig:100micron} but for sSFR$_{\mathrm{MIR}}$. Green dashed vertical line shows the FWHM of the PSF of W4 at the median redshift of our pairs samples.}
\label{fig:W4}
\end{center}
\end{figure*}

\begin{figure*}
\begin{center}

\includegraphics[scale=0.6]{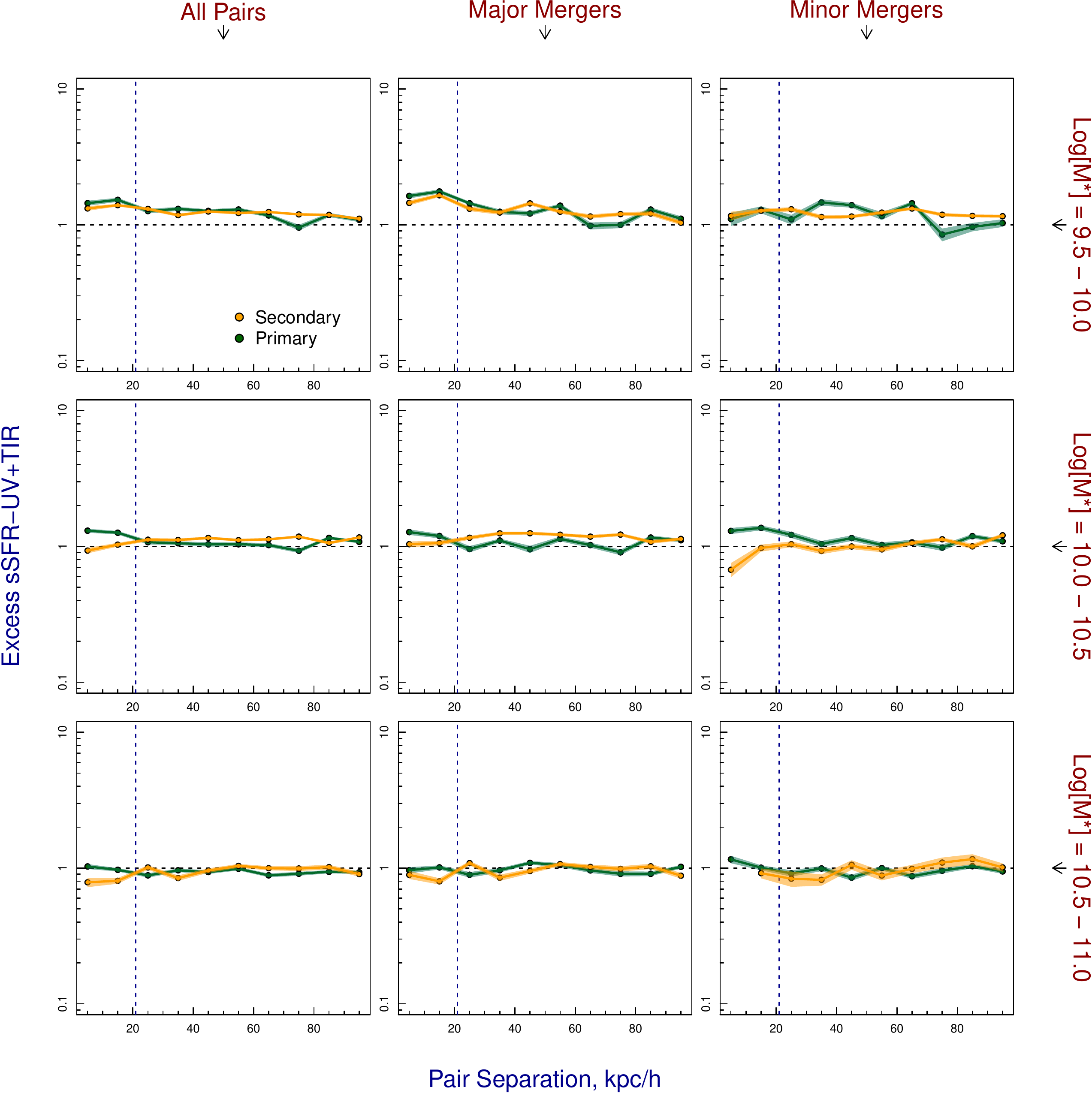}

\caption{Same as Fig. \ref{fig:100micron} but for sSFR$_{\mathrm{UV+TIR}}$. Green dashed vertical line shows the FWHM of the PSF of GALEX-NUV at the median redshift of our pairs samples.}
\label{fig:UV_TIR}
\end{center}
\end{figure*}

\begin{figure*}
\begin{center}

\includegraphics[scale=0.6]{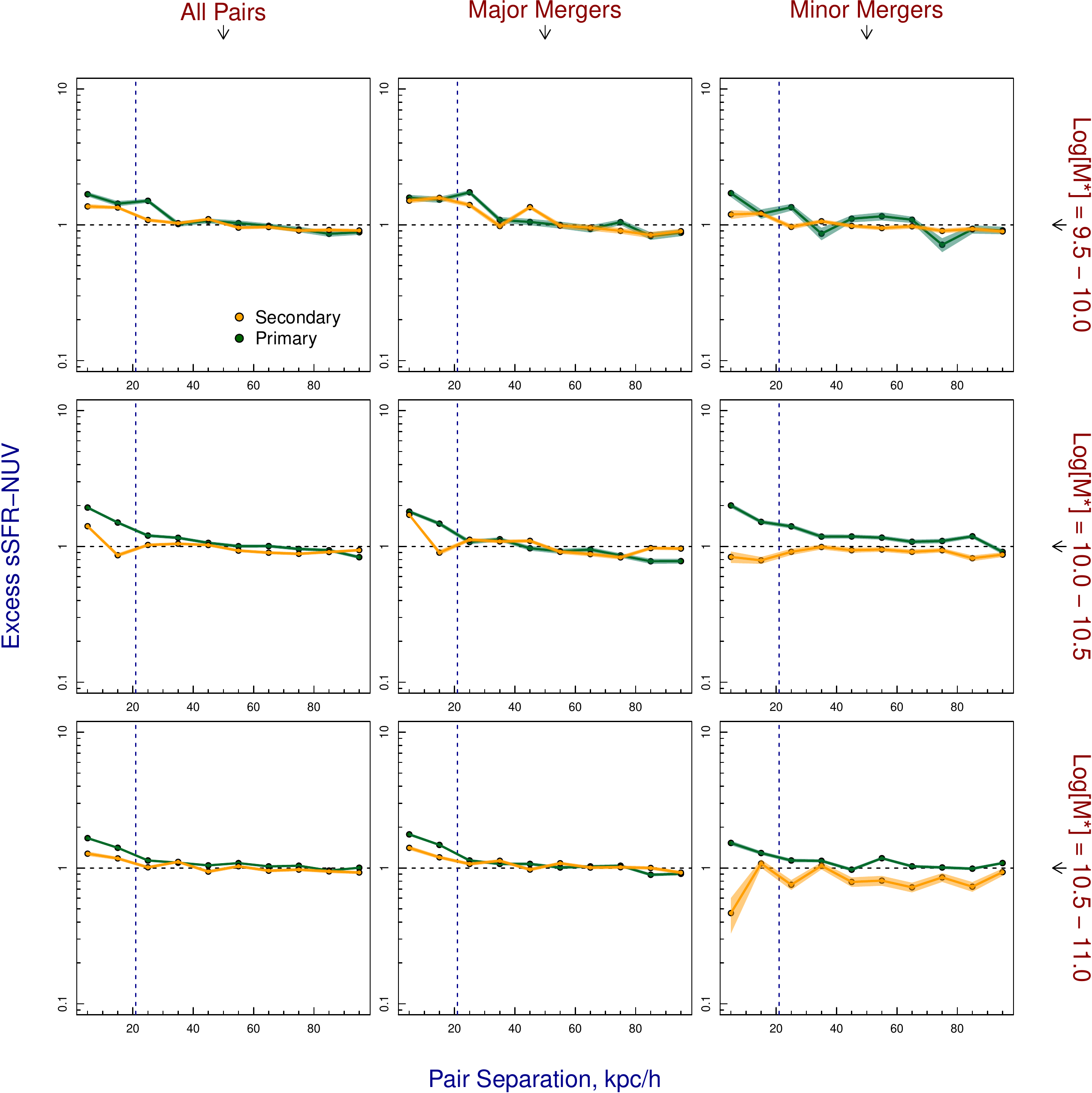}

\caption{Same as Fig. \ref{fig:100micron} but for sSFR$_{\mathrm{NUV}}$. Green dashed vertical line shows the FWHM of the PSF of GALEX-NUV at the median redshift of our pairs samples. For our SFR$_{\mathrm{FUV}}$ indicator the distributions are almost identical with a slight difference in normalisation - as such we do not display both figures.}
\label{fig:NUV}
\end{center}
\end{figure*}

\begin{figure*}
\begin{center}

\includegraphics[scale=0.6]{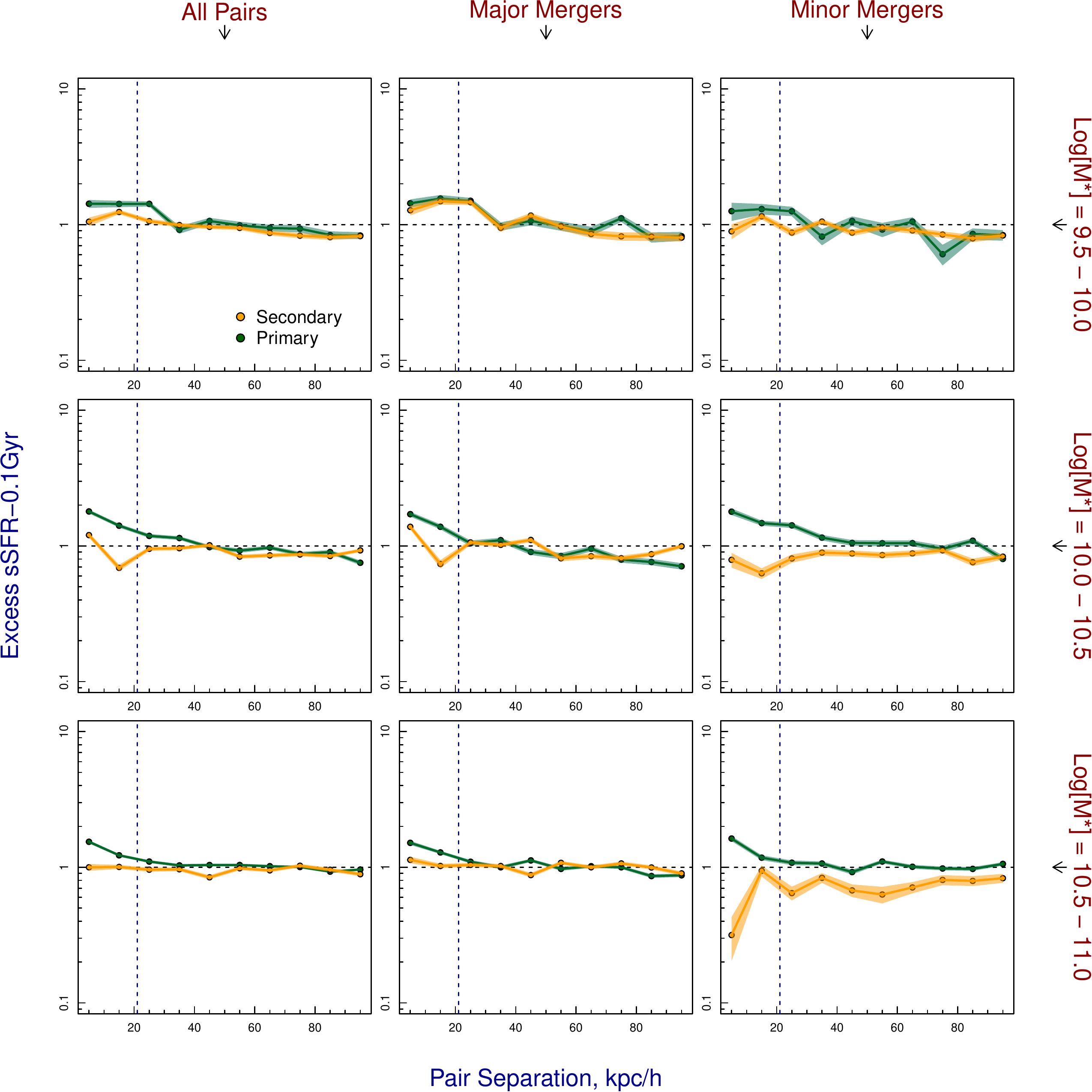}

\caption{Same as Fig. \ref{fig:100micron} but for sSFR$_{\mathrm{0.1Gyr}}$. Green dashed vertical line shows the FWHM of the PSF of GALEX-NUV at the median redshift of our pairs samples - this band is the main driver of SFR calculated using Magphys.}
\label{fig:Magphys}
\end{center}
\end{figure*}

\begin{figure*}
\begin{center}

\includegraphics[scale=0.6]{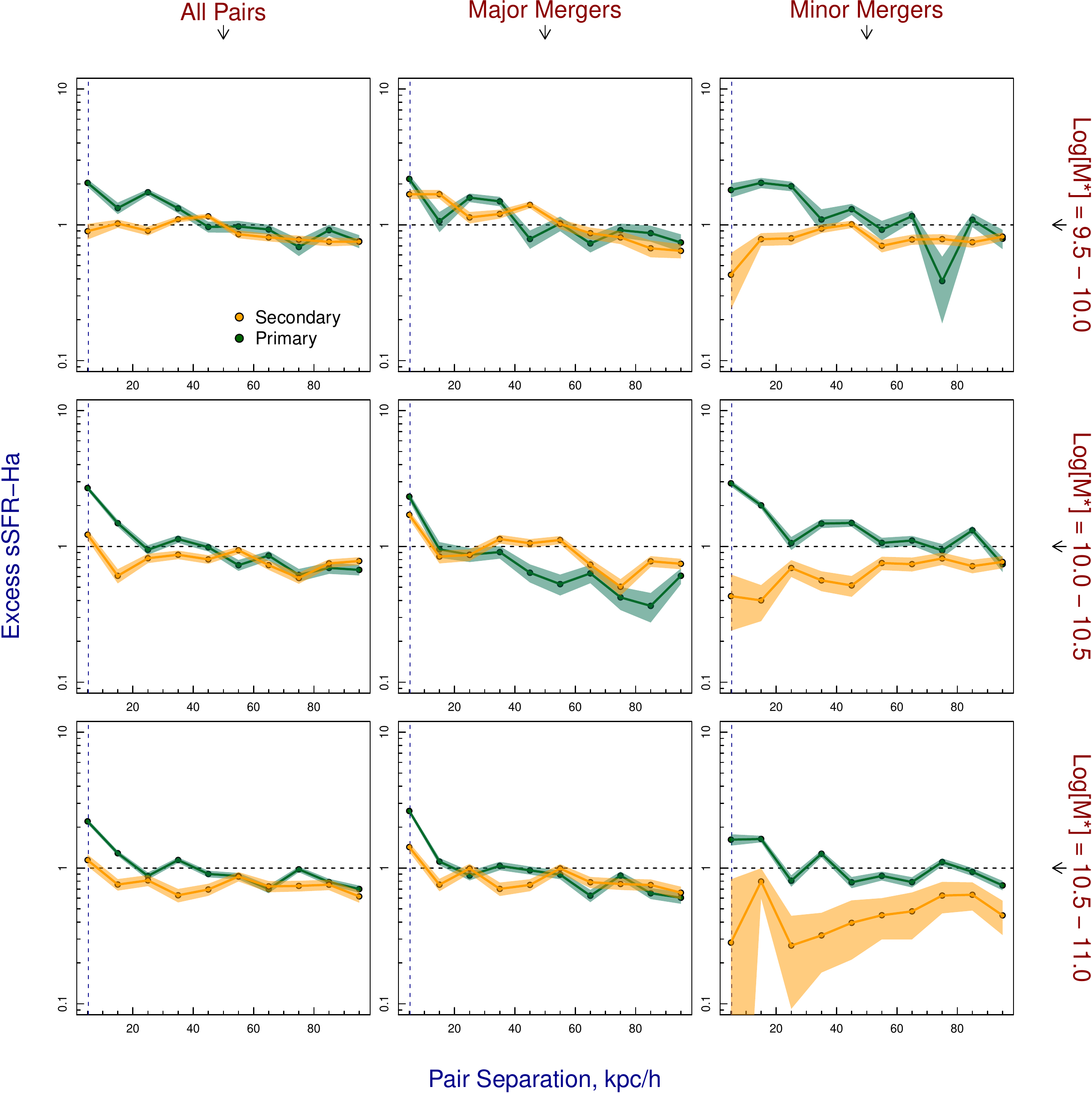}

\caption{Same as Fig. \ref{fig:100micron} but for sSFR$_{\mathrm{H\alpha}}$. Green dashed vertical line shows the physical size of a 2$^{\prime\prime}$ aperture at the median redshift of our pairs samples.}
\label{fig:Ha}
\end{center}
\end{figure*}

\vspace{5mm}

\noindent$\bullet$ \textbf{SFR$_{\mathrm{FIR}}$ (Fig. \ref{fig:100micron})} - Firstly considering our full sample (left column), we find no strong enhancement or suppression of SFR$_{\mathrm{FIR}}$ as a function of pair separation - with both primary and secondary pair galaxies showing no real excess/deficit in comparison to the control sample at all pair separations. This is consistent with the results seen in Fig. \ref{fig:aaron_class}, that there is no significant difference in FIR measure sSFR when considering close-, intermediate- and far-pairs. Splitting this sample into major and minor mergers, we find that the major merger systems (middle column) are almost identical to the full pairs sample and also show no excess/deficit in SF.  

However, in minor merger systems we see some differences between primary and secondary galaxies. Primary galaxies show marginal enhancement of SF as a function of decreasing pair separation (especially in the 10.0$<$log$_{10}$[M*]$<$10.5 range - middle panel of the right column), while secondary galaxies show a deficit in SF in comparison to the control sample. This suppression also appears stronger with increasing stellar mass (from top to bottom of the right column). We also note that, while this suppression is most apparent at very close pair separations, the trend of increasing deficit in SF as a function of pair separation begins outside of the PACS 100$\mu$m PSF (blue vertical dashed line) and as such is not likely to be driven by deblending confusion.

\vspace{5mm}
 \noindent$\bullet$ \textbf{SFR$_{\mathrm{MIR}}$ (Fig. \ref{fig:W4})} - For all samples using our MIR SFR indicator we see marginal enhancement of SF as a function of decreasing pair separation (all samples have a weak slope which increases to small pair separations). We see little difference between primary and secondary galaxies in all samples.

 \vspace{5mm}
 \noindent$\bullet$ \textbf{SFR$_{\mathrm{UV+TIR}}$ (Fig. \ref{fig:UV_TIR})} - For our UV+TIR SFR indicator we find no strong enhancement/suppression of SF with decreasing pair separation. We do see some enhancement in SF for major mergers in our lowest mass bin (top panel of the middle column), and a minor difference between the primary and secondary galaxies in minor mergers at  10.0$<$log$_{10}$[M*]$<$10.5 (middle panel of right column).

 \vspace{5mm}
 \noindent$\bullet$ \textbf{SFR$_{\mathrm{NUV}}$ (Fig. \ref{fig:NUV})} - For the first of our short duration SFR indicators ($<100$\,Myr) we start to see strong trends in our samples as a function of pair separation. For all pair galaxies combined (left column) we find increasing SF with deceasing pair separation for both primary and secondary galaxies. Splitting into major and minor mergers, for major mergers we find that primary and secondary galaxies look identical, both have increasing star-formation with decreasing pair separation. However, for minor mergers we see subtle differences between primary and secondary galaxies. The primary galaxies show a similar relation to those in major mergers, while the secondary galaxies have no enhancement/suppression of SF with pair separation, except for the closest pairs. As with our FIR indicator, there is also a subtle suggestion that the normalisation of the secondary galaxies in minor mergers scales as a function of galaxy mass, with higher mass galaxies appearing more suppressed than lower mass galaxies (the orange line is lower in the bottom panel than the top panel of the right column). We do not display both FUV and NUV figures as both distributions are almost identical, but with slight normalisation scaling.

 \vspace{5mm}
 \noindent$\bullet$ \textbf{SFR$_{\mathrm{0.1Gyr}}$ (Fig. \ref{fig:Magphys})} - Our SFR$_{\mathrm{0.1Gyr}}$ measure displays similar results to the SFR$_{\mathrm{NUV}}$, but also with a slight normalisation change (This is not surprising as they are all derived from the Magphys fits). However, for minor mergers and our highest mass bin (bottom right panel) we now see clear suppression of secondary galaxies at close pair separations.

  \vspace{5mm}
 \noindent$\bullet$ \textbf{SFR$_{\mathrm{H\alpha}}$ (Fig. \ref{fig:Ha})} - Here we see the strongest variation in SFRs as a function of pair separation and the most dramatic differences between primary and secondary galaxies in minor mergers. When considering all pairs (left column) we find an increase in SFR for both primary and secondary galaxies (with the enhancement much more dramatic in primary galaxies). This is once again consistent with the results seen in Fig. \ref{fig:aaron_class}, where H$\alpha$ derived SFRs show a strong enhancement of SF when considering close-, intermediate- and far-pairs. However, when we consider just pair galaxies in minor mergers (right column), we see a dramatic change in the SFR with regards to primary/secondary status. Primary galaxies still have their SF strongly enhanced, while secondary galaxies show strong suppression of SF, even at large pair separations (out to $\sim$r$_{\mathrm{sep}}>$90\,kpc/h).This highlights that SF is suppressed in the secondary galaxies of minor mergers. As SFR$_{\mathrm{H\alpha}}$ measures SFRs on the shortest timescales, it is likely that the enhancement and suppression seen in primary and secondary galaxies respectively is a direct consequence of the galaxy interactions. We remind the reader that our H$\alpha$ flux measurements are not subject to measurement errors from confusion between close pairs, and as such are not systematically biased by errors in flux distribution. 
Another possible explanation for the trends seen in H$\alpha$, are that the distribution of star-forming regions within a galaxy changes during the interaction. The recent simulations of \cite{Moreno15} show that star-formation can be significantly enhanced in the central regions of a galaxy during an interaction, but is largely suppressed in its outer parts. As our H$\alpha$ observations only probe the central regions (due to the aperture-based spectroscopy) we may only be witnessing centrally concentrated suppression/enhancement of star-formation which is not a true representation of the global affects of the interaction. However, the trends seen in our H$\alpha$ SFRs are consistent with those in our other short-duration SFR indicators, albeit at a higher significance. Further investigation into the spatial distribution of H$\alpha$ SF in a subsample of our interacting galaxy sample is underway, and will be the subject of an a upcoming paper.

   \vspace{5mm}
 \noindent$\bullet$ \textbf{Global observation} - An interesting additional global observation (as alluded to previously), is that pair galaxies appear to become more suppressed in SF as a function of increasing stellar mass. This can be seen in all figures in that the lines systematically drop to lower excess sSFR from the top to bottom panels, and suggests that SF in pairs galaxies is not only affected by pair mass ratio and primary/secondary status, but also individual galaxy mass.

 \subsubsection{Summary of Key Observables in this Section}
 \label{sec:summary1}
 
In summary, we find that the primary galaxies in both major and minor mergers show consistent trends of  enhanced SF over our control sample as a function of decreasing pair separation, and that this enhancement is more pronounced in the short duration SFR indicators of SFR$_{\mathrm{H\alpha}}$, SFR$_{\mathrm{FUV/NUV}}$ and SFR$_{\mathrm{0.1Gyr}}$. In contrast, the SF in secondary galaxies shows different characteristics depending on pair mass ratio. In major mergers SF is enhanced at close pair separations, and follows the primary galaxies, while in minor mergers, we see the converse, that secondary galaxies appear to be suppressed at close pair separations. This effect is also more pronounced for short timescale SFR indictors.  We find that at large pair separations (r$_{\mathrm{sep}}>$60\,kpc/h) pair galaxies look similar in their SFR characteristics and as such, we are likely to be probing these systems prior to significant effects of the interaction. However, we see significant modifications to SFRs at close pair separations, suggesting the effect is due to the interaction. We also find that the normalisation of the distributions in Figures \ref{fig:100micron} - \ref{fig:Ha} marginally drops as a function of stellar mass, which potentially indicates that individual galaxy mass also has an impact on how interactions modify SF.

What is clear from this analysis is that both primary and secondary status within the pair and pair mass ratio play an important role in the modification of SFRs in interactions and we most clearly see the effect of these parameters when measuring SFRs on short timescales.

\section{Comparisons of SFR indicators to probe SF on different timescales}

In this section we take our analysis further and combine the effects of both pair separation and SFR indicator timescale. We also use this analysis to highlight the key differences between primary and secondary galaxies, and major and minor mergers discussed above.

\subsection{Scatter in the SFR$_{H\alpha}$ vs SFR$_{FIR}$ relation}
\label{sec:scatter}

Prior to a more sophisticated analysis, and as alluded to earlier, Fig. \ref{fig:SFR_comp} highlights a possible avenue for exploring SFRs in close pairs using multiple SFR indicators. The SFR$_{H\alpha}$ vs SFR$_{FIR}$ relation in Fig. \ref{fig:SFR_comp} displays large scatter with significant offset between H$\alpha$ and FIR derived values for a large number of sources. As the SFR$_{FIR}$ probes a much longer timescale than SFR$_{H\alpha}$, could this highlight short timescale changes in SF for pair galaxies? 

\begin{figure}
\begin{center}

\includegraphics[scale=0.58]{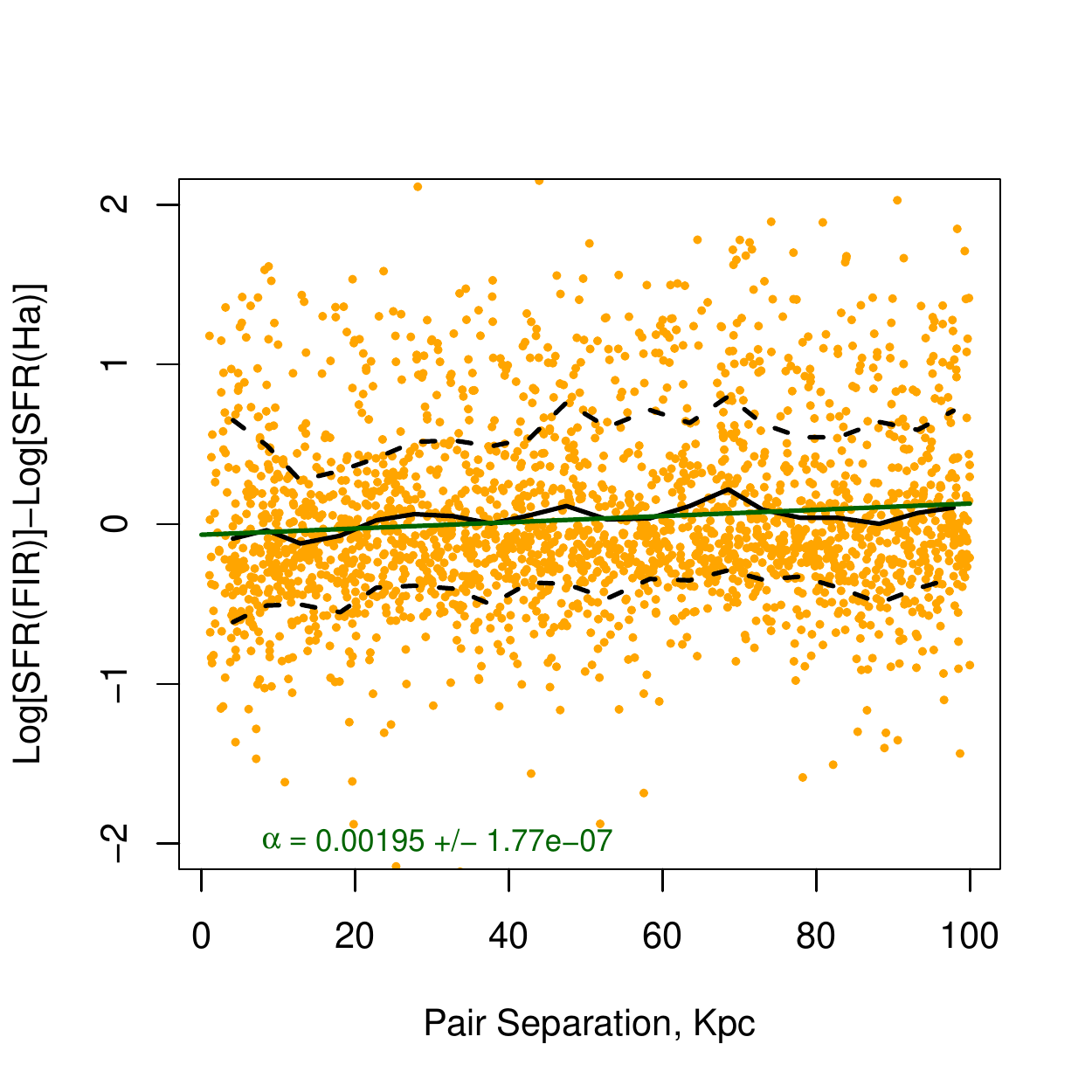}

\caption{log$_{10}$[SFR$_{FIR}$] - log$_{10}$[SFR$_{H\alpha}$] for our pairs sample as a function of pair separation. Black lines show the running median (solid) with upper and lower quartiles (dashed). The green solid line displays the \textsc{Hyperfit} fit to the median binned distribution (with slope, $\alpha$). While marginal we see a slight trend in log$_{10}$[SFR$_{FIR}$] - log$_{10}$[SFR$_{H\alpha}$] suggesting when considering all pair galaxies together, SF may be enhanced on short timescales in closely interacting pairs.}
\label{fig:scatter}
\end{center}
\end{figure}

In Fig. \ref{fig:scatter} we show the offset between log$_{10}$[SFR$_{FIR}$] and log$_{10}$[SFR$_{H\alpha}$] for pair galaxies as a function of pair separation. The black lines in Fig. \ref{fig:scatter} display the running median (solid) and upper and lower quartiles (dashed). The green solid line displays the \textsc{Hyperfit} fit to the median binned values, with slope, $\alpha$, displayed on the plot. While only marginal, we do see a slight trend between log$_{10}$[SFR$_{FIR}$] and log$_{10}$[SFR$_{H\alpha}$] as a function of pair separations. Galaxies at large pair separations have slightly larger log$_{10}$[SFR$_{FIR}$] than log$_{10}$[SFR$_{H\alpha}$] while at small pair separations, the inverse is true. This alludes to the fact the SF, as measured by H$\alpha$ line emission, is very marginally enhanced over that measured from the FIR continuum for close pairs - potentially highlighting that the enhancement of SF occurs on timescales of $<100$\,Myr. However, the Spearman's rank correlation coefficient for log$_{10}$[SFR$_{FIR}$] - log$_{10}$[SFR$_{H\alpha}$] against pair separation if just 0.0929, indicating that there is little statistical correlation between the two variables. We shall investigate this potential correlation using a more sophisticated method in the following section. 

\subsection{Direct comparison of all SFR indicators}
\label{sec:compare}

Taking this further, in this section we compare all of our SFR indicators directly and discuss the implications of variations between each measure. In order to make this comparison we use the slope of the sSFR vs pair separation relations in Figures \ref{fig:100micron} - \ref{fig:Ha} over the full 100\,kpc separation range. We once again use the \textsc{Hyperfit} package to fit the slope of each pair separation binned distribution and give 1$\sigma$ errors. While the distributions in Figures \ref{fig:100micron} - \ref{fig:Ha} may not be well fit by a straight line, we stick to this simple model to avoid complication, and to highlight the general trend of increasing/decreasing excess star-formation as a function of pair separation. 

In Fig. \ref{fig:timescale_change} we show the \textsc{Hyperfit} slope for all SFR indicators split by stellar mass, major/minor mergers and primary/secondary status. The vertical blue dashed line separates long ($>100$\,Myr) and short ($<100$\,Myr) duration SFR indicators, as defined earlier in this work. We also highlight the H$\alpha$ derived SFR slope in the grey shaded region, noting that this point is derived from nebular emission line measurements from aperture spectroscopy, and as such is subject to different biases to our continuum derived SFRs. The dashed horizontal line displays the dividing line between positive and negative slopes, this highlights either enhancement (slope increasing to smaller pair separations) or suppression (slope decreasing to smaller pair separations) of SF in the interaction in comparison the the most distant pairs ($i.e.$ do close pairs show more or less excess SF than distant pairs?). As such, any significant deviation from zero slope displays that the interaction is affecting SF for that particular sample. We note here again, that we also fit these distributions outside of the PSF scale for the main photometric band used in the analysis (blue vertical dashed lines in Figures \ref{fig:100micron} - \ref{fig:Ha}), and find that the correlations hold true - but at a less significant level. 

Comparing the slopes of the distribution as a function of SFR indicator, we see some interesting trends. Firstly, considering just galaxies in major mergers, we find that as we move to SFR indicators that probe shorter timescales, star-formation is enhanced more strongly (for long duration measures there is no star-formation enhancement, while for short duration measures we see consistent enhancement of star-formation). This is true for both primary and secondary galaxies. It is also interesting to note that once again there appears to be a weak trend with stellar mass, however, this is only apparent in secondary galaxies, with secondary galaxies being less enhanced at increasing stellar mass (the orange points drop from the top to bottom panels). In contrast for galaxies in minor mergers, we find that primary and secondary galaxies show different characteristics. Primary galaxies display similar trends to the primary galaxies in major mergers, showing excess star-formation for short duration measures. Secondary galaxies largely show suppression of star-formation, in all but the smallest mass bin. What is clear from this figure is that for short duration SFR indicators, secondary galaxies in minor mergers show different characteristics to their primary counterparts.

These results are indicative of SF being enhanced in primary galaxies of all mergers and secondary galaxies being enhanced in major mergers and suppressed in minor mergers, but only when measured on short timescales.  

To highlight the dependance on primary/secondary status and pair mass ratio of the modification of SFRs in interactions further, we compare differences between the slopes for major and minor mergers, at fixed stellar mass binning and pair status (Fig. \ref{fig:timescale_change_MM}) and differences between primary and secondary galaxies at fixed pair mass ratio (Fig. \ref{fig:timescale_change_PS}). Error bars in these figures are derived from the sum of the squares of the \textsc{Hyperfit} errors given in Fig. \ref{fig:timescale_change}. In Fig. \ref{fig:timescale_change_MM}, offsets from the central line display a significant difference in the effect of interactions between major and minor mergers, where positive values show that SF is more strongly suppressed in a minor merger. Note that this figure does not show absolute values and as such zero does not highlight that there is no change to SFRs in the interaction, only that there is  \textit{no difference} between major/minor mergers  galaxies. We find that primary galaxies are largely consistent with their being no difference between major and minor mergers for all SFR indicators (the green points are largely consistent with the central line given the errors), while the secondary galaxies stronger suppression of SF in minor mergers than major mergers (the orange points sit above the central line). This effect appears largest in short duration SFR indicators ($i.e.$ the differences between secondary galaxies in major and minor mergers is most apparent when probed on short timescales).

In Fig. \ref{fig:timescale_change_PS} offsets from the central line display a significant difference in the effect of interactions between primary and secondary galaxies, where positive values show that SF is more strongly suppressed in the secondary galaxy. We find that for major mergers there is little difference between primary and secondary galaxies specifically for long duration indicators (as discussed previously). We do see that for short duration timescale indicators, secondary galaxies are marginally suppressed in comparison to primary galaxies when considering the higher stellar mass bins only (the gold and purple triangles sit above the central line), but this effect is small. For minor mergers we see a larger discrepancy between primary and secondary galaxies. Secondary galaxies appear suppressed relative to primaries for all star-formation rate indicators and in the majority of stellar mass bins (the majority of points in the right panel of Fig. \ref{fig:timescale_change_PS} sit above the central line). This effect also appears more pronounced for short duration indicators and at intermediate stellar masses.      

In combination, these figures highlight that the secondary galaxies in minor mergers show distinctly different SF characteristics to all other pair galaxies, but this effect is only strong when SFRs are measured on $<100$\,Myr timescales.

\section{Summary of the effect of close interactions on galaxy SFRs and discussion}
\label{sec:summary2}

Now we outline the key results outlined in the previous two sections and discuss a possible toy model for SF in galaxy mergers which may explain these results:
 \vspace{2mm}

\noindent$\bullet$ Considering pair populations as a whole, SF can appear either unaffected or enhanced by an interaction depending on SFR indicator used (Fig. \ref{fig:aaron_class}). 
 \vspace{2mm}
 
 \noindent$\bullet$ At large pair separations ($>$50\,kpc) we see little difference between different galaxy populations, suggesting these systems are not being strongly effected by the interaction (Figures \ref{fig:100micron} - \ref{fig:Ha}).
 
  \vspace{2mm}
 
 \noindent$\bullet$ At close pair separations ($<$30\,kpc) we see both enhancement and suppression of galaxy SFRs depending on the subsample of galaxies we select, and this modification to SFRs appears more strongly for short duration SFR indicators (Figures \ref{fig:100micron} - \ref{fig:Ha}).

 \vspace{2mm}
  
\begin{figure*}
\begin{center}

\includegraphics[scale=0.8]{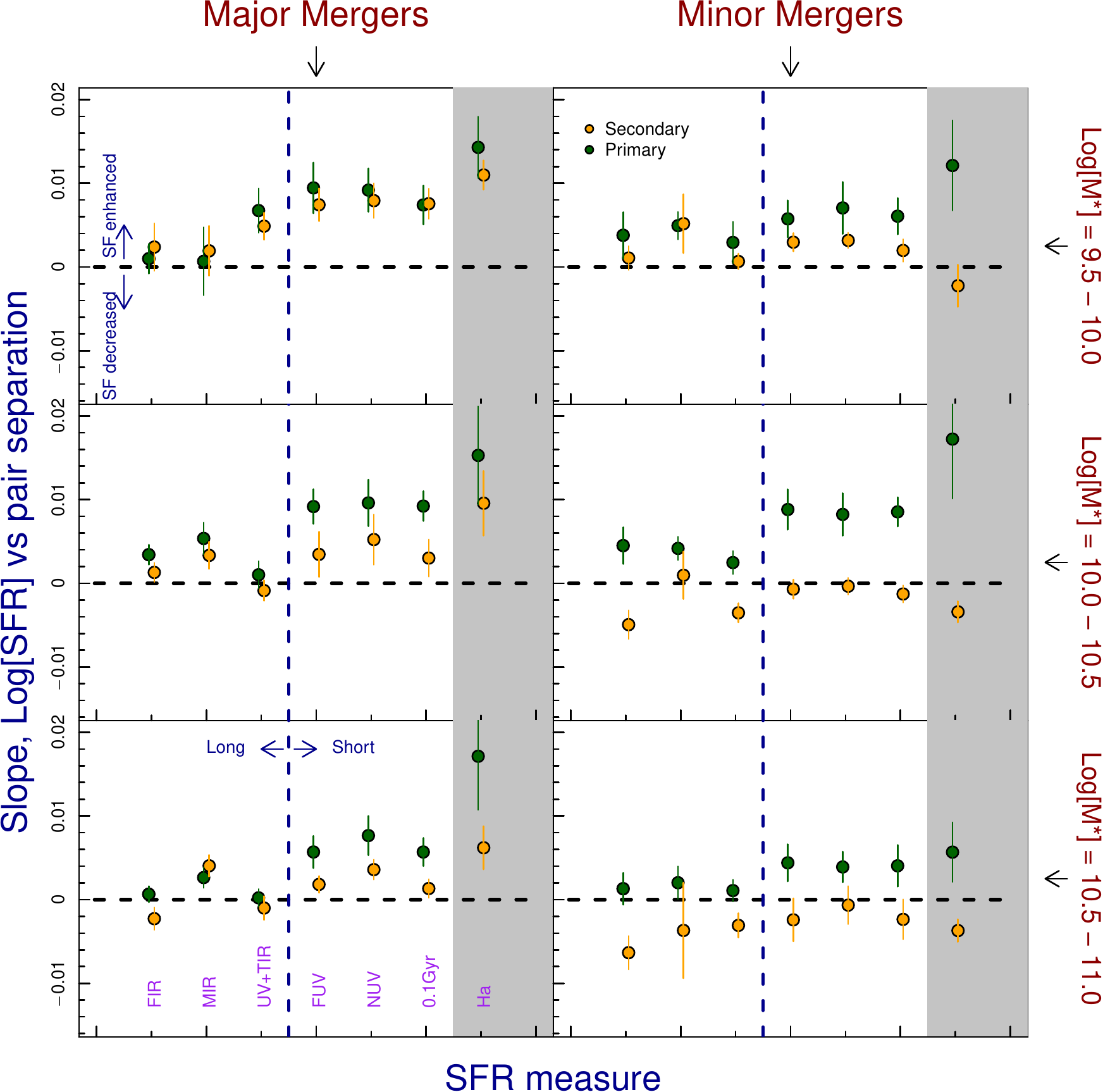}

\caption{Comparison of the slopes of log$_{10}$[sSFR] vs pair separation for all SFR indicators shown in Figure \ref{fig:100micron} - \ref{fig:Ha}. Samples are split on stellar mass (rows) and pair mass ratio (columns), as well as primary (green) and secondary (orange) status within the pair. We divide of SFR indicators into long and short duration with the vertical blue dashed line, and highlight the H$\alpha$ emission line derived SFR with the grey shaded region. Significant deviations from the central horizontal line display modifications to SF induced by the a merger - positive values display enhancement of SF and negative values suppression of SF.  Error bars display the 1$\sigma$ error on the hyper.fit fit to the slope.  }
\label{fig:timescale_change}
\end{center}
\end{figure*}

\noindent$\bullet$ For long timescale SFR indicators, we see little enhancement or suppression of SF in interacting galaxies. This holds true for both major and minor mergers, and for both primary and secondary galaxies - suggesting that we are probing the SF in the galaxy prior to its SFR being modified by the interaction (Fig. \ref{fig:timescale_change}).    
 \vspace{2mm}
 
\noindent$\bullet$ For short timescale SFR indicators (especially H$\alpha$), primary galaxies show enhancement in SF at close pair separations, which is consistent across major and minor mergers (Fig. \ref{fig:timescale_change}).
 \vspace{2mm}
 
\begin{figure*}
\begin{center}

\includegraphics[scale=0.7]{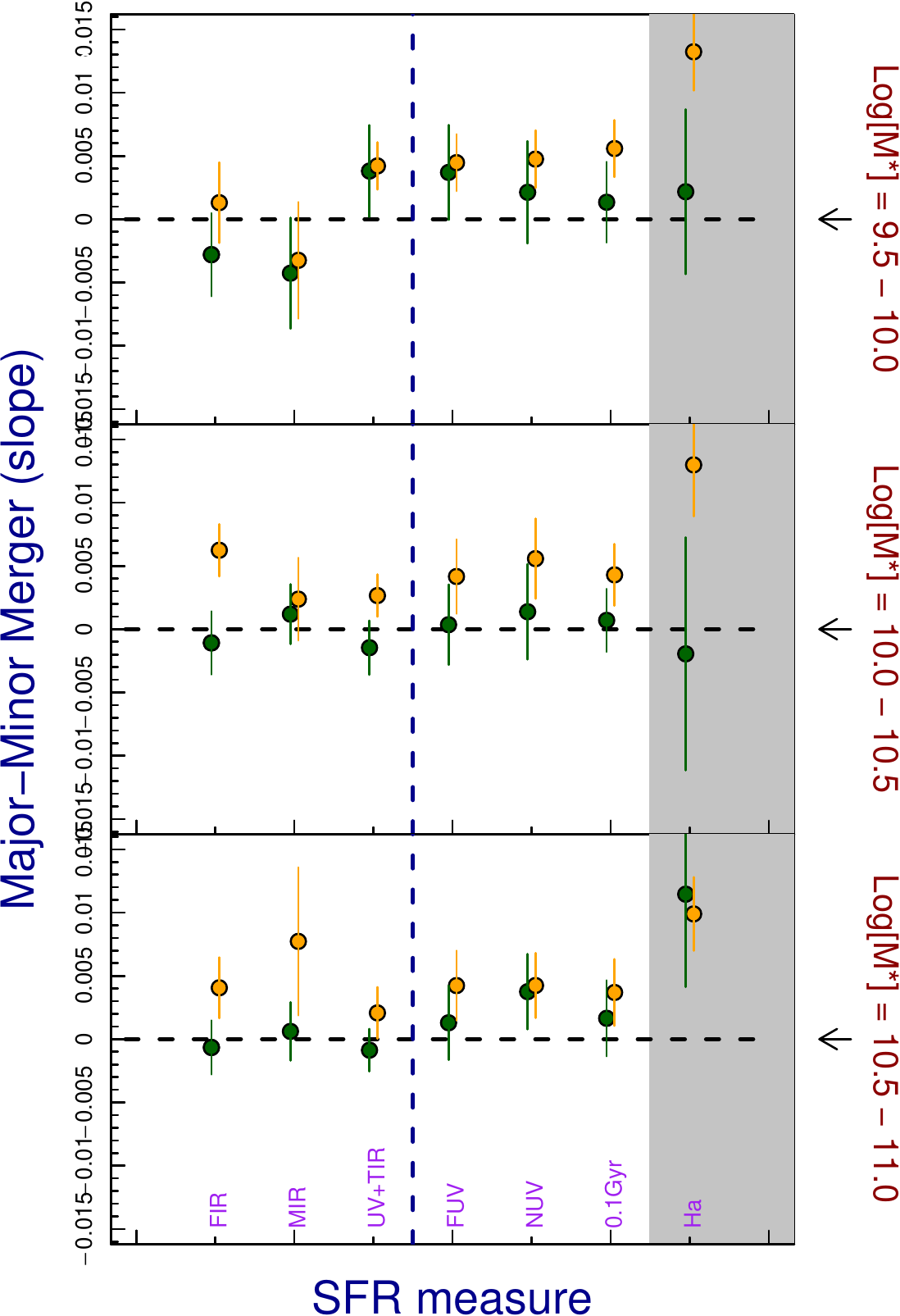}

\caption{Comparison of the difference of the log$_{10}$[sSFR] vs pair separation slope between major and minor mergers. Values given are the offset between corresponding points in the left and right column in Fig. \ref{fig:timescale_change}. Significant offsets from the central line show differences between major and minor mergers for each population and SFR indicator. Once again, secondary galaxies in show more significant differences between major and minor mergers.   }
\label{fig:timescale_change_MM}
\end{center}
\end{figure*}

\begin{figure*}
\begin{center}

\includegraphics[scale=0.8]{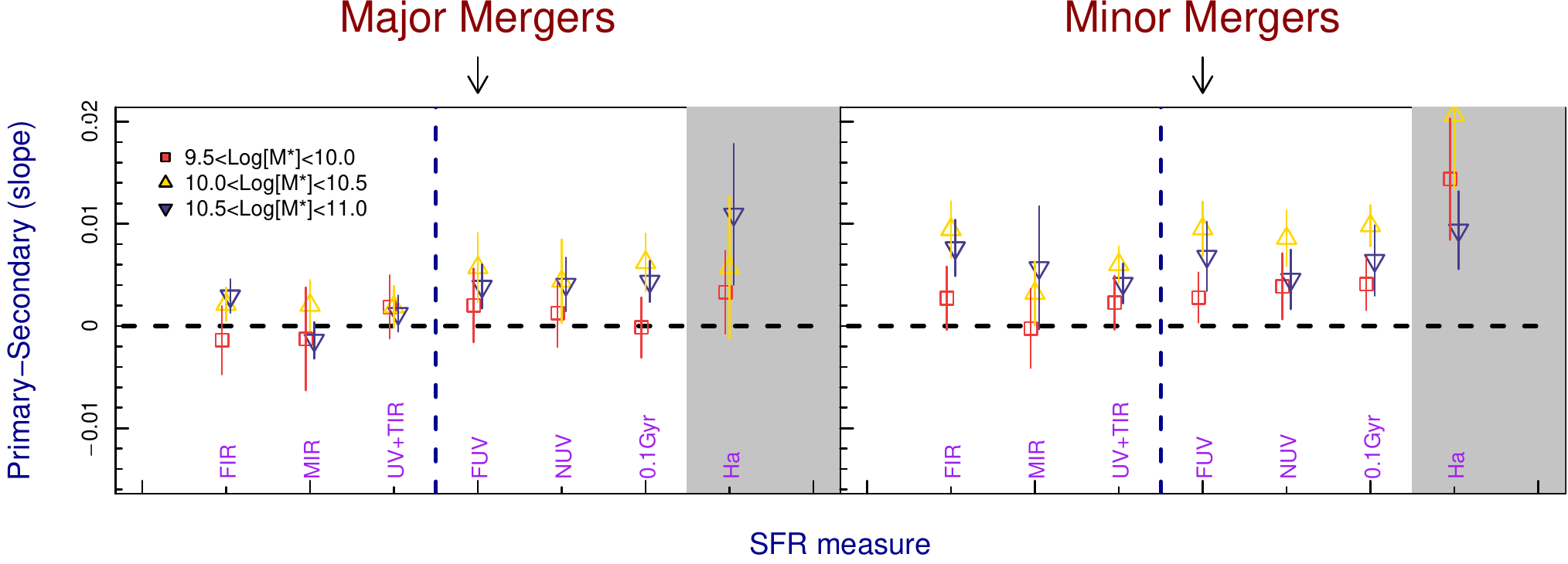}

\caption{Comparison of the difference of the log$_{10}$[sSFR] vs pair separation slope between primary and secondary galaxies. Values given are the offset between primary and secondary galaxies in each row in Fig. \ref{fig:timescale_change}. Significant offsets from the central line show differences between primary and secondary galaxies at each stellar mass and SFR indicator. Once again, secondary galaxies in minor mergers in show more significant differences between primary and secondary galaxies, specifically for short duration SFR indicators.  }
\label{fig:timescale_change_PS}
\end{center}
\end{figure*}

\noindent$\bullet$ For short timescale SFR indicators, secondary galaxies show distinctly different characteristics between major and minor mergers. In major mergers SF is enhanced with decreasing pair separation, while in minor mergers SF is suppressed with decreasing pair separation (Figures \ref{fig:timescale_change}, \ref{fig:timescale_change_MM} \& \ref{fig:timescale_change_PS}).
 \vspace{2mm}
 
\noindent$\bullet$  Primary galaxies appear largely agnostic to pair mass ratio while secondary galaxies show suppression in minor mergers only (Figures \ref{fig:timescale_change}, \ref{fig:timescale_change_MM} \& \ref{fig:timescale_change_PS}).

  \vspace{2mm}

To relate these observables to the true effects of interactions on SFRs, we use both the pair separation and variation of SFR indicators on different timescales as a proxy for the stage of the galaxy interaction. We relate both our long duration SFR indicators at all separations and our short duration measures at large pair separations to be representative of the pre-, or early-interaction stage. Conversely, our short duration measures at close pair separations is representative of the late-interaction stage. This simplistic model assumes that galaxies are all on their first approach and that they have not had previous passes, which would effect their star-formation. Such systems could reach relatively large pair separations, but would have already had their star-formation enhanced/suppressed by the interaction. In our analysis we have no method of determining if galaxies have already had one or more passes prior to the observation epoch, as such we can not account for this affect. However, we note that \cite{Robotham14} find that, for the pair sample used in this work, galaxies at $r_{sep}<20$\,kpc show much greater signs of visual disturbance than those at $20<r_{sep}<50$\,kpc and both these populations show higher visual disturbance than pairs at $r_{sep}>50$\,kpc. This indicates that galaxies are more strongly affected by interactions at closer separations, and while a fraction of galaxies at large separations my have already had a number of passes, the major trend is that closer separations mean stronger pair interactions and a later stage in the merger process. Therefore, we deem it is reasonable to assume that to first order pair separation can be directly related to merger timescale.

\begin{figure*}
\begin{center}

\includegraphics[scale=0.42]{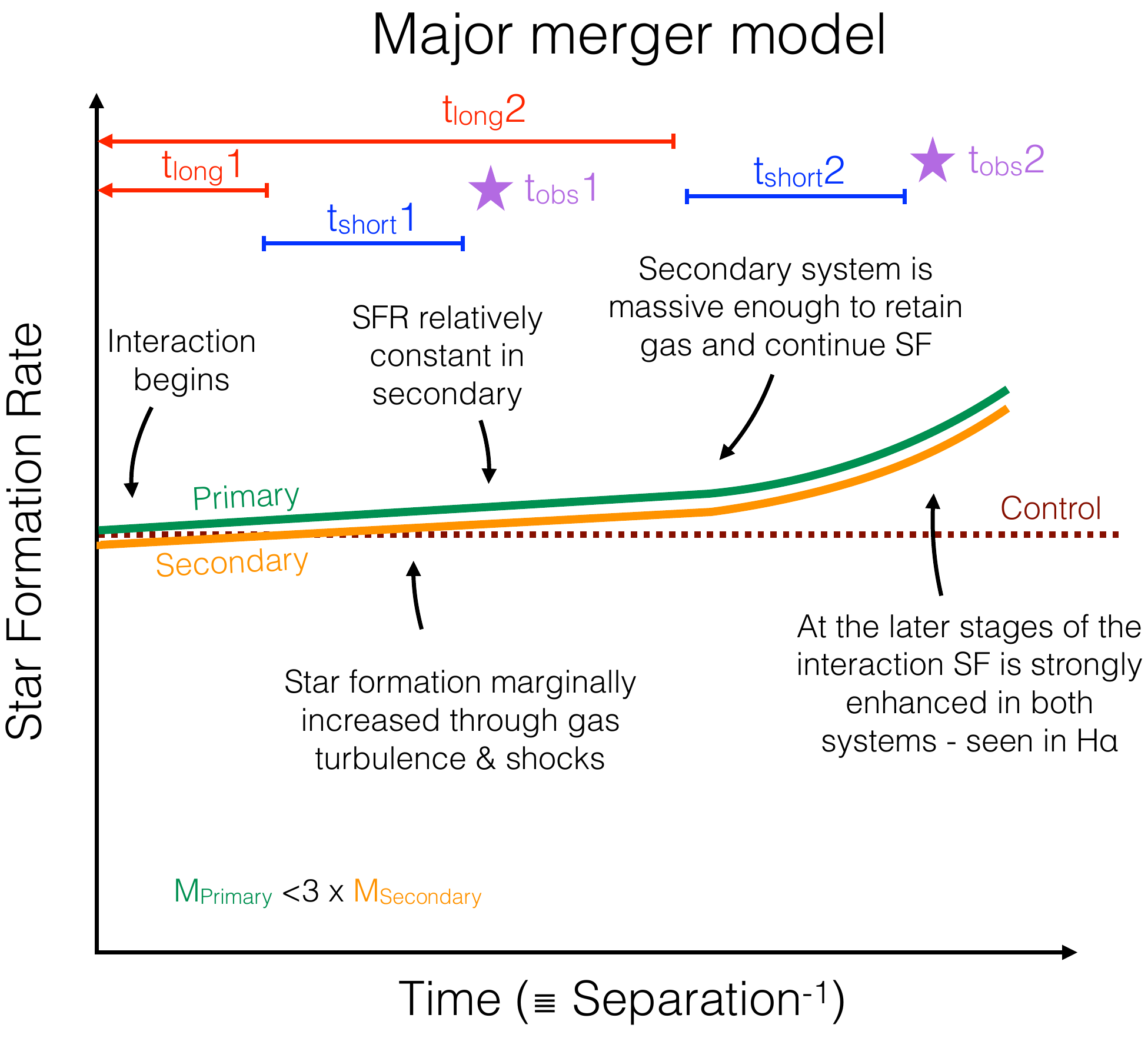}
\includegraphics[scale=0.42]{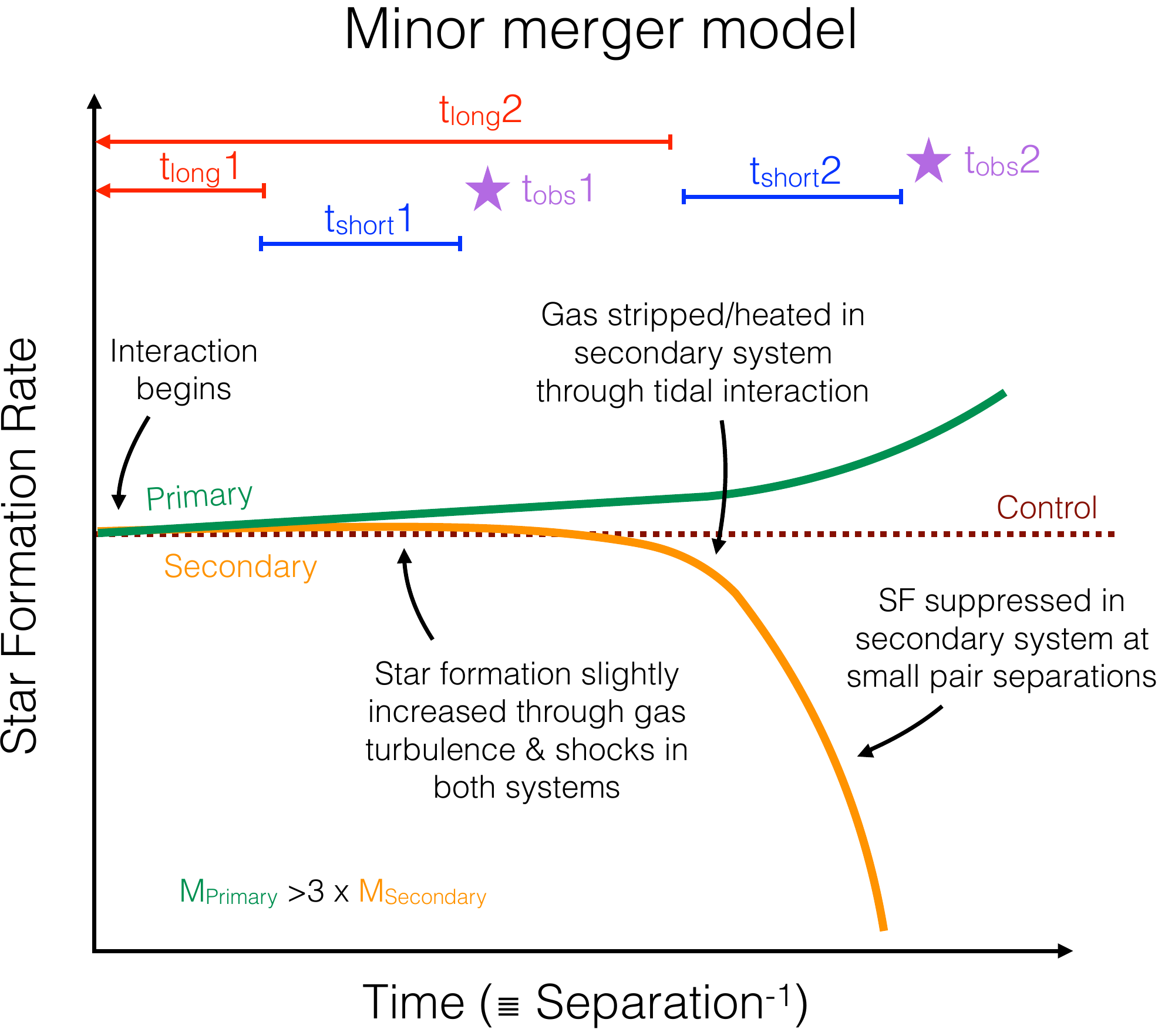}

\caption{A toy model for the evolution of SF in major (left) and minor (right) merger systems. The purple stars highlight two possible observation epochs during the merger. These are proceeded by the likely timescales over which long duration and short duration SFRs are measured at each epoch (red and blue horizontal bars respectively - note red bars would extend off the left edge of the plot). These models are consistent with our observations of SF in pair systems and can potentially explain the seemingly contradictory previously obtained results for SF in close pairs as SF is enhanced in all merger scenarios, except in the secondary galaxy in minor mergers at close separation. As such, our measure of how SF is affected in galaxy interaction is dependent on pair mass ratio and primary/secondary status within the pair.}
\label{fig:cartoon_model}
\end{center}
\end{figure*}

Using these assumptions we can piece together a hypothetical toy model for SF in the primary and secondary galaxies in both major and minor merger systems.   Fig. \ref{fig:cartoon_model} displays this hypothetical model with interaction time/pair separation plotted against current SFR. We also display the timescales over which both long and short duration SFRs are measured, for two different scenarios: t1, when the galaxies are at large pair separations (an early stage in their interaction) and t2, when the galaxies are at small pair separations (a late stage in their interaction). 

In our major merger model SF remains relatively constant throughout the merger in both the primary galaxy and secondary galaxy, but is enhanced at the very late stages (we see the strongest enhancement in the shortest timescale indicator - H$\alpha$). These enhancements in SF are likely to be be due to gas turbulence, tidal torques or shocks cause by the gravitational attraction of the other system. In this scenario, the secondary galaxy has sufficient mass (in comparison to the primary) to retain its gas and continue star-forming at a relatively constant rate $i.e.$ it is not tidally stripped. This is consistent with our observations summarised above. The primary  and secondary galaxies show some enhancement of SF with pair separation, and this increases for the shortest timescale indicator. If SFRs were measured at t1 and t2 we would observe a marginally increasing measure of both long and short duration SFRs (the primary and secondary galaxy show similar SF properties which increase in SF for short duration measures). The largest tidal interaction will occur at the smallest pair separation and be most evident in the shortest duration SFR indicator, as such we see the largest enhancement of SF in H$\alpha$ derived SFRs.  

For our minor merger model, the primary galaxy follows the same SF evolution as both systems in the major merger. It is sufficiently high mass in comparison to the secondary to retain its gas and continue star-forming, and has its SFR enhanced over time as interactions with the secondary systems cause gas turbulence and shocks. However, the secondary system follows a different SF evolution path during the interaction. Initially its SF is not significantly altered and hence we do not see distinct differences in long duration SFR measures or a large pair separations, which probe this epoch. However, at some stage during the interaction SF begins to be suppressed. This is likely through tidal stripping or gas heating. This is evident as the suppressed SFRs at close pair separations and in short duration SFR indicators. If we were to measure the SFR of the secondary system at t1 and t2, we would find that at t1 (large separation), the galaxy would look somewhat like the primary galaxy, and we would see no suppression of SF. However, at t2 we would find short duration SFRs suppressed relative to t1, this effect would not yet be observable in the long duration SFR measures. At this point the secondary galaxy SFRs would be distinctly different from the primary galaxy when observed using short duration indicators - this model is also consistent with the observations summarised above.  

If representative of the true galaxy-galaxy interactions, these toy models could potentially explain the contradictory results from previous studies which find both SF enhancement and suppression in close pair systems. Our models predict that SF is enhanced in the majority of cases - where mass ratios are close to 1:1, as well as in the primary galaxy in a minor merger. However, SF is suppressed in the secondary galaxy in minor mergers, but only at close pair separations. As such, whether SF is enhanced or suppressed in galaxy-galaxy interactions is largely dependent on the pair mass ratio and pair separation. Clearly by observing galaxies at different interaction stages and at different pair mass ratios, we will will obtain conflicting results as to the suppression/enhancement of SF in interacting systems.

\section{Conclusions}

We have investigated star-formation activity as a function of pair separation for interacting galaxies for different merger scenarios, using SFRs derived from multiple SFR indicators covering a range of timescales. The key results from our analysis are as follows:
\\

\noindent $\bullet$ SF in interactions proceeds differently for galaxies in minor mergers than in major mergers.\\
\\
\noindent $\bullet$ SF in minor mergers proceeds differently in galaxies depending on whether they are the primary (high mass) or secondary (low mass) galaxy.\\
\\
\noindent $\bullet$ Primary galaxies show some enhancement of SF in all mergers which is only evident the shortest timescales - and as such the later stages of the interaction. \\
\\
\noindent $\bullet$ Secondary galaxies show little change in SF in major mergers, except for some enhancement on the shortest timescales, but are suppressed in minor mergers. \\
\\
\noindent $\bullet$ SF suppression in secondary galaxies occurs on timescales of $\lesssim100$\,Myr as it is not evident in long duration SFR measures - as such, it also occurs at a late stage in the interaction.\\

Using these results we propose a scenario for the evolution of SF in major and minor merger systems. In the major mergers both primary and secondary galaxies have their SF marginally enhanced through tidal turbulence and shocks for the early stages. At the later stages of the merger SF is likely to be enhanced. SF is not suppressed as both galaxies have sufficient mass (in relation to the other) to retain their gas. In the minor mergers the primary galaxy follows the same SF enhancement as the major merger systems, while the secondary galaxy has its SF suppressed as its gas is either tidally stripped, heated or simply stretched to a lower mean density which is not sufficient for SF. This toy model is consistent with the recent findings of \cite{Robotham13} and \cite{DePropris14}, as well as the galaxy merger simulation of \cite{DiMatteo07} - predicting that the effects of galaxy interactions on star-formation are dependant on both pair mass ratio and primary/secondary status within the pair.

\section*{Acknowledgements}

GAMA is a joint European-Australasian project based around a spectroscopic campaign using the Anglo-Australian Telescope. The GAMA input catalogue is based on data taken from the Sloan Digital Sky Survey and the UKIRT Infrared Deep Sky Survey. Complementary imaging of the GAMA regions is being obtained by a number of independent survey programs including GALEX MIS, VST KiDS, VISTA VIKING, WISE, Herschel-ATLAS, GMRT and ASKAP providing UV to radio coverage. GAMA is funded by the STFC (UK), the ARC (Australia), the AAO, and the participating institutions. The GAMA website is http://www.gama-survey.org/. MJIB acknowledges financial support from the Australian Research Council (FT100100280). NB acknowledges support from the EC FP7 SPACE project ASTRODEEP (Ref. no. 312725). L. Dunne acknowledges support from European Research Council Advanced Investigator grant COSMICSM.

\appendix
 
\section{Robotham et al Sample Split on Major/Minor Merger and Primary/Secondary Status}
 
\begin{figure*}
\begin{center}

\includegraphics[scale=0.6]{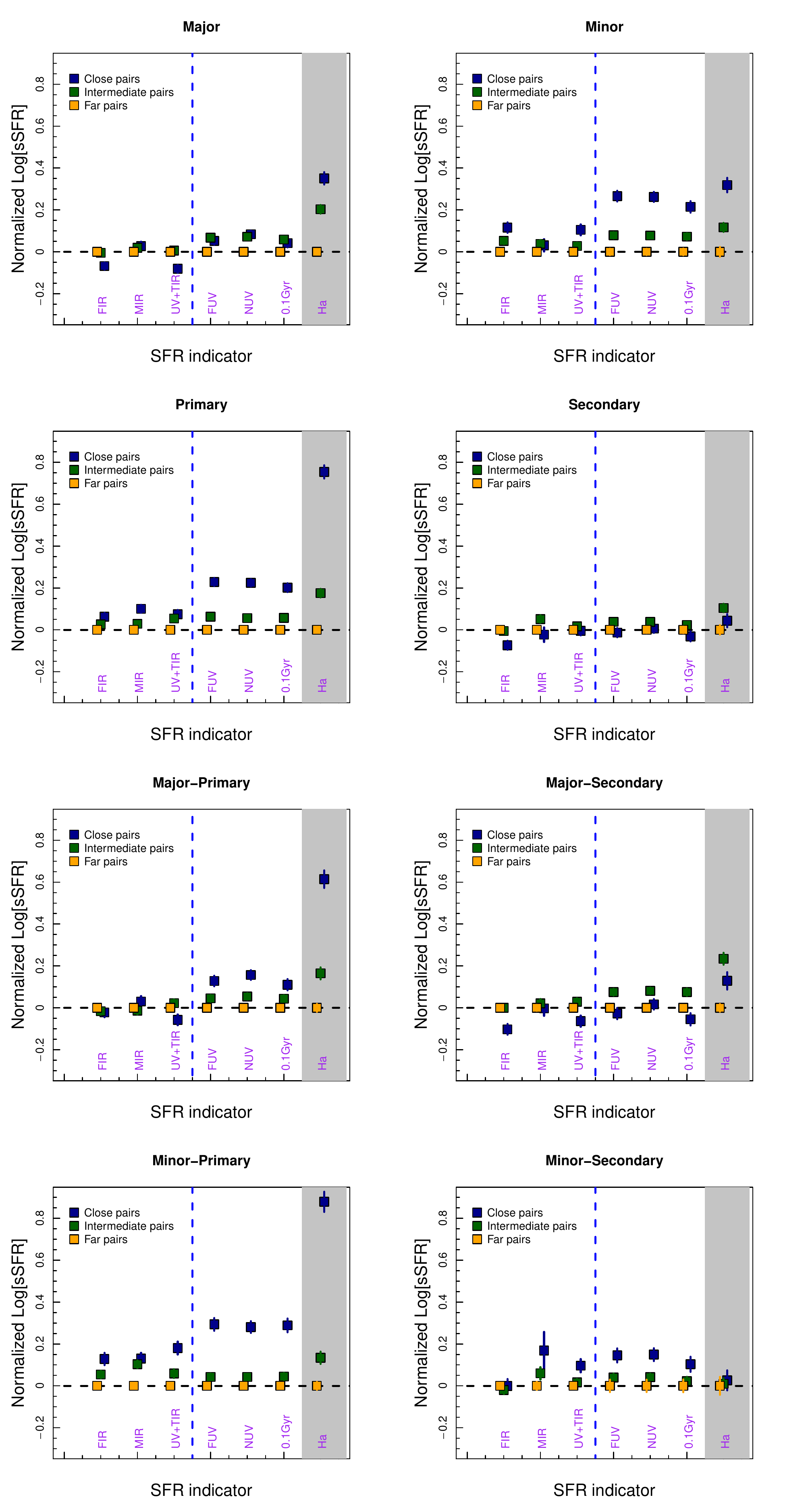}

\caption{Pair galaxies selected from \citet{Robotham14} split by major/minor mergers and primary/secondary status. All pair samples are normalised to the far-pairs' value to highlight the effects of interactions on pair galaxies in each SFR indicator. }
\label{fig:arron_all}
\end{center}
\end{figure*}

In Fig. \ref{fig:arron_all} we display the pair sample of  \cite{Robotham14} discussed in Sec. \ref{sec:SF_pairs} but split into major/minor mergers and primary/secondary status. These distributions echo the main results derived in this paper. Considering the populations as a whole, we see enhancement of star-formation in close pairs, but this enhancement is only observable in SFR indicators which probe short timescales. Splitting the samples into primary and secondary galaxies, we find  that the increase in SF for close pairs is completely driven by the primary galaxies, where strong excess is seen for short duration indicators (specifically H$\alpha$). Splitting further on both pair mass ration and galaxy status, we find that it is the primary galaxies in both major and minor mergers that show strong enhancement in SF. For secondary galaxies we see little difference between far-, intermediate- and close-pairs for all SFR indicators. This is consistent with the main results in this paper, when derived in a separate, but complementary, manner.

\end{document}